\documentclass[twocolumn]{aastex631}
\usepackage{tabularx}
\usepackage{xspace}
\newcommand{\prot}{\ensuremath{P_{\mbox{\scriptsize rot}}}}
\newcommand{\bprp}{\ensuremath{G_{\rm BP} - G_{\rm RP}}}

\newcommand{\alphafe}{$[\mathrm{\alpha}/\mathrm{Fe}]$}

\newcommand{\teff}{$T_{\rm eff}$}
\newcommand{\ith}{\ensuremath{^{\rm th}}}

\newcommand{\halpha}{high-$\alpha$}
\newcommand{\logg}{$\log g$\xspace}
\newcommand{\vsini}{$v$sin$i$}
\newcommand{\Mdot}{$M_{\odot}$}

\begin{document}

\title{Spectroscopic Follow-up of Young High-$\alpha$ Dwarf Star Candidates: Still Likely Genuinely Young}

\correspondingauthor{Yuxi(Lucy) Lu}
\email{lucylulu12311@gmail.com}

\newcommand{\amnh}{American Museum of Natural History, Central Park West, Manhattan, NY, USA}
\newcommand{\cca}{Center for Computational Astrophysics, Flatiron Institute, 162 5\ith\ Avenue, Manhattan, NY, USA}
\newcommand{\columbia}{Department of Astronomy, Columbia University, 550 West 120\ith\ Street, New York, NY, USA}
\newcommand{\anu}{Research School of Astronomy $\&$ Astrophysics, Australian National University, Canberra ACT 2611, Australia}
\newcommand{\osu}{Department of Astronomy, The Ohio State University, Columbus, 140 W 18th Ave, OH 43210, USA}
\newcommand{\astrothreed}{Center of Excellence for Astrophysics in Three Dimensions (ASTRO-3D), Australia}
\newcommand{\ccapp}{Center for Cosmology and Astroparticle Physics (CCAPP), The Ohio State University, 191 W. Woodruff Ave., Columbus, OH 43210, USA}
\newcommand{\utah}{Department of Physics \& Astronomy, University of Utah, Salt Lake City, UT 84112, USA}

\author[0000-0003-4769-3273]{Yuxi(Lucy) Lu}
\affiliation{\osu}
\affiliation{\ccapp}

\author[0000-0002-0900-6076]{Catherine Manea}
\affiliation{\utah}

\author[0000-0001-6180-8482]{Maryum Sayeed}
\affiliation{\columbia}

\author[0000-0001-7371-2832]{Stephanie T. Douglas}
\affiliation{Lafayette College, 730 High St, Easton, PA 18042, USA}

\author[0000-0002-6502-1406]{Madeleine McKenzie}
\affiliation{Carnegie Observatories, 813 Santa Barbara St, Pasadena, CA, 91101, USA}

\author[0000-0003-2431-981X]{Dominick Rowan}
\affiliation{\osu}
\affiliation{\ccapp}
\affiliation{Department of Astronomy, University of California Berkeley, Berkeley, CA 94720, USA}

\author[0000-0002-0551-046X]{Ilya Ilyin}
\affiliation{Leibniz-Institute for Astrophysics Potsdam (AIP), An der Sternwarte 16, D-14482, Potsdam, Germany}

\author[0000-0001-7339-5136]{Adam Wheeler}
\affiliation{Center for Computational Astrophysics, Flatiron Institute, 162 5th Avenue, New York, NY 10010, USA}

\author[0000-0002-4031-8553]{Sven Buder}
\affiliation{\anu}
\affiliation{\astrothreed}

\author[0000-0003-2508-6093]{Louis Amard}
\affiliation{Departemnt of Astronomy, University of Geneva, Chemin Pegasi 51, CH-1290 Versoix, Switzerland}

\author[0000-0002-7549-7766]{Marc H. Pinsonneault}
\affiliation{\osu}
\affiliation{\ccapp}

\author{Jennifer A. Johnson}
\affiliation{\osu}
\affiliation{\ccapp}

%% Note that the \and command from previous versions of AASTeX is now
%% depreciated in this version as it is no longer necessary. AASTeX 
%% automatically takes care of all commas and "and"s between authors names.

%% AASTeX 6.31 has the new \collaboration and \nocollaboration commands to
%% provide the collaboration status of a group of authors. These commands 
%% can be used either before or after the list of corresponding authors. The
%% argument for \collaboration is the collaboration identifier. Authors are
%% encouraged to surround collaboration identifiers with ()s. The 
%% \nocollaboration command takes no argument and exists to indicate that
%% the nearby authors are not part of surrounding collaborations.

%% Mark off the abstract in the ``abstract'' environment. 
\begin{abstract}
The question of whether genuinely young \halpha\ stars exist has been discussed for over a decade since their discovery from asteroseismology of giant stars as it is challenging to break the degeneracy between the binary interaction and the genuinely young scenarios.
Young \halpha\ stars are hard to explain with traditional chemical evolution model as the \halpha\ disk is typically associated with the early epoch of star formation in the Milky Way.
Combined with recent advances of gyrochronology, and that $^7$Li can serve as an unambiguous indicator for identifying merger products in dwarfs thanks to its low burning temperature, we identified young \halpha\ dwarf candidate stars through their fast rotation in a previous study. 
In this paper, we performed high-resolution spectroscopic follow-up of these candidates using Potsdam Echelle Polarimetric and Spectroscopic Instrument (PEPSI), and confirm 3 additional stars that are most likely genuinely young.
Together with the star from the earlier paper, we find three out of four of them center around [Fe/H]=$-0.5$ dex, are $\sim$ 5 Gyr old, and have a similar amount of elevated Li ($\sim$0.5 dex) and Al ($\sim$0.1 dex) compared to stars with matching \logg, \teff, Mg, and Fe within observational uncertainties, hinting at their common formation pathway.
\end{abstract}

%% Keywords should appear after the \end{abstract} command. 
%% The AAS Journals now uses Unified Astronomy Thesaurus concepts:
%% https://astrothesaurus.org
%% You will be asked to selected these concepts during the submission process
%% but this old "keyword" functionality is maintained in case authors want
%% to include these concepts in their preprints.

%% From the front matter, we move on to the body of the paper.
%% Sections are demarcated by \section and \subsection, respectively.
%% Observe the use of the LaTeX \label
%% command after the \subsection to give a symbolic KEY to the
%% subsection for cross-referencing in a \ref command.
%% You can use LaTeX's \ref and \label commands to keep track of
%% cross-references to sections, equations, tables, and figures.
%% That way, if you change the order of any elements, LaTeX will
%% automatically renumber them.
%%
%% We recommend that authors also use the natbib \citep
%% and \citet commands to identify citations.  The citations are
%% tied to the reference list via symbolic KEYs. The KEY corresponds
%% to the KEY in the \bibitem in the reference list below. 

\section{Introduction} \label{sec:intro}
The discovery of stars with enhanced $\alpha$-element abundances yet unexpectedly young ages presents a significant challenge to conventional models of Galactic chemical evolution \citep{Chiappini2015, Martig2015, SilvaAguirre2018, Claytor2020, Das2020, Zinn2022, Grisoni2024, Lu2025}. 
These so-called ``young high-$\alpha$ stars'' are particularly puzzling because \alphafe\ is positively correlated with the ratio of enrichment from Type II Supernova (produced by exploding massive stars) to enrichment from Type Ia Supernova (produced by white dwarf mergers or accretions).
Stars with enhanced \alphafe\ are typically associated with rapid early star formation in the Milky Way, where star formation efficiency is high \citep[e.g.,][]{Conroy2022}, implying ages greater than $\sim$8 Gyr. 
Yet, mounting evidence mostly from asteroseismology \citep[e.g.,][]{Chiappini2015, Martig2015, SilvaAguirre2018, Zinn2022, Grisoni2024, Jofre2023}, and from stellar rotation \citep{Claytor2020, Lu2025} has revealed a small but robust population of $\alpha$-enhanced stars that appear significantly younger.
The creation of these young \halpha\ stars could be possible but typically requires rapid local enrichment without phase mixing the star-forming gas \citep[e.g.,][]{Johnson2021, Garver2023}.

One major difficulty in interpreting these stars is disentangling true youth from the effects of binary evolution. 
Asteroseismology measures the current mass of the star, which is related to age under the assumption that the star has experienced no significant interactions during its lifetime. 
As a result, the asteroseismic candidates are more precisely defined as massive $\alpha$-rich stars, which are young for a normal stellar evolution history.
However, mass transfer, mergers, or tidal spin-up in close binaries can all conspire to obscure a star's true age, either by increasing its mass or by altering its surface properties \citep[e.g.,][]{Tayar2015, Yong2016, Jofre2016, Hekker2019, Zhang2021, Jofre2023, Cerqui2023, Grisoni2024, Izzard2018}. 
For red giants, stellar mergers are difficult to distinguish from genuinely young stars.
Merger products spin down quickly as stars expand, so they need not be rapid rotators. 
Most chemical species are not impacted if both stars are on, or close to, the main sequence.

Lithium is an interesting exception to this general guideline.
Li provides direct evidence for distinguishing between merger and genuinely young scenarios: it is easily destroyed during main-sequence star mergers \citep[e.g.,][]{Ryan2001, Pinsonneault2002, Ryan2002}, as the thin convective envelope and limited Li reservoir means Li can be quickly depleted by heating associated with mass transfers.
The relatively shallow convection zones of dwarf stars mean that Li is less diluted compared to in evolved stars, making it easier to detect.

By leveraging gyrochronology as an age indicator for dwarf stars \citep{Barnes2003}, \cite{Lu2025} discovered one likely true young \halpha\ star and a list of similar candidates using Kepler \citep{Borucki2010} and K2 \citep{Howell2014} for rotation period measurements, and APOGEE \citep{Majewski2017} and GALAH \citep{DeSilva2015} for abundances.
To exclude binary interactions, not only do we need Li detections, but close-in binaries also need to be excluded, as they can tidally spin-up a star without any mass transfer \citep[e.g.,][]{Meibom2005, Lurie2017, Simonian2019}.
Unfortunately, most of these candidates identified in \cite{Lu2025} only have data from GALAH, which lacks multi-epoch radial velocity (RV) needed to exclude close-in binaries, or are only observed in APOGEE, which lacks Li measurements. 

In this paper, we present new high-resolution spectroscopic follow-up of some of the young \halpha\ dwarf star candidates from \cite{Lu2025} to investigate their chemical abundances, binary information, kinematics, and Li content in detail. 
Our goal is to further test whether any \halpha\ dwarf stars are genuinely young and their possible formation pathways. 
The target selection is described in Section~\ref{subsec:targets}.
Analysis procedure including  reduction (Section~\ref{subsec:reduction}), RV measurements and corrections (Section~\ref{subsec:rv_measure}), abundance measurement (Section~\ref{subsec:abund_analysis}), period measurement (Section~\ref{subsec:period}), initial candidate exclusion (Section~\ref{subsec:initial_exclude}), binary analysis (Section~\ref{subsec:binary_analy}), and comparisons with doppelgänger stars with similar global properties in spectra Signal-to-Noise (SNR), Mg or $\alpha$, Fe, \logg, and \teff (Section~\ref{subsec:abund_doppel}) are shown in Section~\ref{sec:dataselection}.
Using these results, we analyze each of the three genuinely young \halpha\ stars in Section~\ref{subsec:truly young}.
In Section~\ref{subsec:abund_comp}, we examine the chemical patterns of all four genuinely young \halpha.
We also highlight one possible merger product in Section~\ref{subsec:merger}.
We discuss their possible formation pathways in Section~\ref{subsec:formation}.
Finally, we summarize our findings and discuss future directions in Section~\ref{sec:concl}.

\section{Data \& Selection} \label{sec:dataselection}
\subsection{Target \& Filter Selections}\label{subsec:targets}
We use the Potsdam Echelle Polarimetric and Spectroscopic Instrument \citep[PEPSI,][]{Strassmeier2015}, mounted on the the 11.8 m Large Binocular Telescope \citep[LBT; e.g.,][]{LBT2012} in Arizona to perform spectroscopic follow-up.
We choose its lowest spectral resolution of $R \sim$ 50,000 to minimize the exposure time.

We selected 5 stars from the GALAH--K2 young \halpha\ candidate sample in Table 1 of \cite{Lu2025} for radial velocity follow-up and 11 stars from the APOGEE--Kepler/K2 young \halpha\ candidate sample to obtain Li abundances. 
The stars in \cite{Lu2025} Table 1 are high in [$\alpha$/Fe] from APOGEE or GALAH. 
The definitions for \halpha\ in \cite{Lu2025} are \alphafe $>-$0.6[Fe/H]+0.20 for APOGEE, and \alphafe $>-$0.3[Fe/H]+0.19 for GALAH.
Young stars rotate more rapidly than old ones, so we select stars with rotation below $<\sim$ 35 days, or $<\sim$ 8 Gyr, which is the typical age of the \halpha\ disk.
However, synchronized binaries exist with period less than 15 days, so our best candidates have periods between 15 and 35 days. 
Besides all the requirements mentioned above, all the APOGEE stars are selected to have more than two RV measurements agreeing within 1 km/s to vet for close-by stellar binaries that can spin-up the stars, and the GALAH stars are also selected to have Li measurements from \cite{Wang2024}, since we want to use Li as a mass-transfer diagnostic.
Finally, we can use \vsini\ as an alternative measurement of rotation and synchronized binaries, and thus, we also remove stars with \vsini\ $>$ 10 km/s.

For the spectroscopic follow-up, we selected relatively bright stars with Gaia $G$-band magnitudes $\lesssim 14\,\mathrm{mag}$ to minimize the exposure time, and were observable from the LBT for the October 2024 to December 2024 observing season. 
We selected cross dispersions settings CD3 (wavelength range of 480.0-544.1 nm) and CD5 (wavelength range of 627.8-741.9 nm) to capture important abundance information, including reliable $\alpha$ element lines (Mg, Ca, Si) and the Li 6707~\AA\ line.
We want to obtain Li for stars only in APOGEE as APOGEE observes in the infrared and does not include the Li line.
The Li measurements from PEPSI will also serve as an independent confirmation of Li for stars in GALAH.
In the end, we were able to obtain spectra for 9 APOGEE--Kepler/K2 young \halpha\ candidate stars and 4 GALAH--K2 young \halpha\ candidate stars.
Their IDs, $V$ magnitudes, number of observations, representative SNR values close to the Li line calculated for the combined spectra, observation days, and exposure times are shown in Table~\ref{tab:1}.

\begin{table*}[]
    \centering
    \begin{tabular}{c|ccccccc}
         Sample&2MASS ID&Gaia DR3 ID & $V$ mag & $N$ Obs. & Comb. SNR & Obs. Date & Exp. Time [s]\\
         \hline
          A--K2 & 03413908+2359163 & 68164639977312384 & 13.9 & 2 & 103 & Oct. 25, 2024 & 1440  \\
          ... & ... & ... & ... & ... & ... & Nov. 1, 2024 & 1440  \\
          
          G--K2 & 03435417+1719250 & 44497171153467008 & 11 &7 & 132 & Sep. 20, 2024 & 240  \\
          ... & ... & ... & ... & ... & ... & Oct. 8, 2024 & 160  \\
          ... & ... & ... & ... & ... & ... & Oct. 8, 2024 & 160  \\
          ... & ... & ... & ... & ... & ... & Oct. 23, 2024 & 120  \\
          ... & ... & ... & ... & ... & ... & Oct. 31, 2024 & 120  \\
          ... & ... & ... & ... & ... & ... & Oct. 31, 2024 & 120  \\
          ... & ... & ... & ... & ... & ... & Nov. 1, 2024 & 240  \\
          
          G--K2 & 01084954-0030464 & 2533196969884293888 & 13.6& 3 & 82 & Oct. 21, 2024 & 1200  \\
          ... & ... & ... & ... & ... & ... & Oct. 31, 2024 & 1200  \\
          ... & ... & ... & ... & ... & ... & Nov. 24, 2024 & 600  \\
          
          \hline
          A--K & 18432994+4356589 & 2117326900207932288 & 14.2&1& 47 & Sep. 20, 2024 & 1800  \\
          
          \hline
         G--K2 & 01064189-0027182 & 2533233700444556800 & 13.8&5 & 111 & Sep. 22, 2024 & 1320   \\
         ... & ... & ... & ... & ... & ... & Oct. 10, 2024 & 1780  \\
         ... & ... & ... & ... & ... & ... & Oct. 10, 2024 & 1780  \\
         ... & ... & ... & ... & ... & ... & Oct. 23, 2024 & 1320  \\
         ... & ... & ... & ... & ... & ... & Oct. 31, 2024 & 1320  \\
         
         G--K2 & 08280198+1913343 & 662897729347117312 & 13.0&4 & 85 & Oct. 8, 2024 & 800   \\
         ... & ... & ... & ... & ... & ... & Oct. 8, 2024 & 800  \\
         ... & ... & ... & ... & ... & ... & Oct. 25, 2024 & 600  \\
         ... & ... & ... & ... & ... & ... & Nov. 1, 2024 & 600  \\
         A--K2 & 01231989+0031088 & 2534334689540935040 & 13.5&1 & 71 & Oct. 25, 2024 & 1200 \\
         A--K2 & 01244193+0124528 & 2559177811173201920 & 12.9&5 & 92 & Oct. 8, 2024 & 780  \\
         ... & ... & ... & ... & ... & ... & Oct. 8, 2024 & 780  \\
         ... & ... & ... & ... & ... & ... & Oct. 21, 2024 & 600  \\
         ... & ... & ... & ... & ... & ... & Oct. 31, 2024 & 600  \\
         ... & ... & ... & ... & ... & ... & Oct. 31, 2024 & 600  \\
         A--K2 & 04314519+2502357 & 150907970912262016 & 13.7&1 & 117 & Sep. 20, 2024 & 1320   \\
         A--K2 & 08443258+1116500 & 601835320304886400 & 13.7&1 & 50 & Nov. 24, 2024 & 1020   \\
         A--K & 19101724+4944353 & 2132794829787857792 & 10.3&1 & 35 & Sep. 19, 2024 & 30  \\
         A--K & 19380704+4708008 & 2128545109969957888 & 12.3&1& 27 & Oct. 25, 2024 & 1200  \\
         A--K & 19580796+4052023 & 2075066724134857344 & 14.3&1 & 35 & Sep. 20, 2024 & 2100   \\
    \end{tabular}
    \caption{Sample name (G--K2: GALAH--K2; A--K2: APOGEE--K2; A--K: APOGEE--Kepler), 2MASS ID, Gaia DR3 ID, $V$ magnitudes, number of observations, representative SNR values using wavelength range 6688-6695 \AA\ calculated for the combined spectra, observation dates, and exposure times for the sample of observed stars.
    The first three stars are the genuinely young rotators (see Section~\ref{subsec:truly young}), and the fourth star is the possible merger product (see Section~\ref{subsec:merger}).}
    \label{tab:1}
\end{table*}

\subsection{PEPSI Spectra Reduction}\label{subsec:reduction}
The PEPSI 2D echelle spectra are reduced following the procedure described in \citet{Strassmeier18}. 
The standard reduction include
bias over-scan detection and subtraction, scattered light surface extraction and subtraction, definition of échelle orders, weighted extraction of spectral orders, wavelength calibration, and a self-consistent continuum fit to the full 2D image of extracted orders.
For each spectrum, Least-Squares Deconvolution (LSD) is also performed in CD3 using PHOENIX model spectra \citep{PHOENIX} that are close in \teff\ and \logg\ to vet for spectroscopic binaries.

\subsection{Radial Velocity Measurements \& Corrections}\label{subsec:rv_measure}
We measure radial velocities (RVs) with iSpec \citep{BlancoCuaresma14}, which uses 1D cross-correlation  with the synthetic Solar template provided with the installation. We use a wavelength range of 4760 -- 5500 \AA\ to measure the RV, and the RV uncertainties are determined following \citet{Zucker2003}. 
The typical uncertainty associated with our analysis is on the order of 0.05 km/s.

\subsection{Abundance Measurements}\label{subsec:abund_analysis}
For stars with more than one observation, we first corrected the RV shift for the individual spectrum.
The final spectrum used to obtain abundances were then obtained from co-adding the individual spectra by taking the mean.

For these stars, we are mostly interested in their $\alpha$ (Mg, Ca, Si) and Li abundances from PEPSI to independently confirm their high $\alpha$ and investigate their Li abundances.  However, we alsodetermine elemental abundances of other species (C, Al, Sc, Ti, V, Cr, Mn, Co, Ni, Cu, Zn, Y, Zr, Ba, Ce, Nd) to obtain a full chemical profile for each star using our higher resolution (R$\sim$50,000) spectra and validate survey abundances.
%To measure their abundances, we used \texttt{Korg}, a fast 1D LTE spectral synthesis code \citep{korg}.
%We adopted the log$g$ and \teff\ values from APOGEE DR17 or GALAH DR3 catalogs as inputs to \texttt{Korg} to increase efficiency.
%We then fit the rest of the global parameters, namely metallicity ([M/H]), microturbulence ($v$ mic), and rotation broadening ($v$sin$i$).
%Finally, we fit for individual abundances in Mg, Fe, and Li while fixing the global parameters.
%The line list is taken to be the same as GALAH DR3 \citep{Galahdr3}, as it includes the Li lines and various iron lines around it.
To determine stellar abundances, we must first assume a set of stellar parameters ($\rm T_{eff}$, log g, microturbulence (v$_{\rm micro}$), metallicity).  We adopt $\rm T_{eff}$ and log g from either APOGEE or GALAH where available due to their superior precision (MAYBE WE CAN MENTION THE TEFF AND LOGG PRECISION HERE?).  We then use Brussels Automatic Code for Characterizing High accUracy Spectra \citep[BACCHUS,][]{BACCHUS} to determine v$_{\rm micro}$, spectral broadening (which includes instrumental broadening and $vsini$), and individual elemental abundances for up to 20 elements (excluding Li, which is treated separately) from the reduced and continuum normalized PEPSI spectra.  BACCHUS is a spectral synthesis and fitting code designed for high-resolution spectra.  It uses Turbospectrum \citep{Plez2012} to synthesize spectra and determines line-by-line abundances using four distinct fitting methods.  We point readers to  \citet{Hayes2022} and \citet{Manea2025} for a detailed description of BACCHUS and its fitting methods.  We interpolate over the MARCS model atmosphere grid \citep{Gustafsson2008} with BACCHUS for our stellar atmospheric models, and we assume one-dimensional local thermodynamic equilibrium (1D LTE) throughout our analysis.  For our atomic data, we use version 5 of the \textit{Gaia}-ESO linelist  \citep{Heiter2021} and combine molecular transition data from numerous sources: CH from \citealt{Masseron2014}, C2, CN, OH, and MgH from T. Masseron, private communication, SiH from \citealt{Kurucz1992}, and TiO, FeH, and ZrO from B. Pelz, private communication.  For our line selection, we begin with lines that \citet{Heiter2021} indicates have reliable log gf values (\texttt{gfflag} = Y) and are unblended (or only blended with lines of the same species) in the Solar and Arcturus spectrum (\texttt{synflag} = Y).  We then only retain lines that return consistent (within 0.1 dex) abundances across the four BACCHUS abundance determination methods.  This discards lines that are saturated or otherwise poorly modeled by the synthetic spectra.  Our final elemental abundances are determined by taking average abundances from all reliable lines of the same species.  For elements with multiple lines, the line-to-line abundance dispersion is adopted as our uncertainty, and for elements with just one line, we assume an uncertainty of 0.1 dex.  Again, this excludes Li, which we treat separately and discuss in a later paragraph.  We present our line-by-line abundances in Table \ref{tab:bacchus_line_by_line_abunds}, which are reported in [X/H] form assuming solar abundances from \citet{Grevesse2007}.

\begin{table*}[]
    \centering
    \begin{tabular}{cc|cccccc}
Species & Wavelength & 2MASSJ03435417 & 2MASSJ01064189 & ... & 2MASSJ08280198 & 2MASSJ19380704 & Sun \\
\hline
C \scriptsize{I} & 5052.1 &  &  & ... & -0.53 &  & 0.04 \\
C \scriptsize{I} & 5380.3 &  & -0.09 & ... &  &  &  \\
C \scriptsize{I} & 6587.6 &  &  & ... &  &  & 0.09 \\
C \scriptsize{I} & 7111.5 &  &  & ... &  &  &  \\
C \scriptsize{I} & 7113.2 &  & -0.24 & ... &  &  & 0.01 \\
$[$O \scriptsize{I} $]$ & 6300.3 &  &  & ... &  & 0.43 & 0.14 \\
Al \scriptsize{I} & 6696.0 & -0.97 & -0.33 & ... & -0.58 & -0.44 & 0.09 \\
Al \scriptsize{I} & 6698.7 &  &  & ... &  &  & 0.1 \\
Mg \scriptsize{I} & 6318.7 & -0.66 & -0.19 & ... &  & -0.13 & 0.24 \\
Mg \scriptsize{I} & 6319.2 &  &  & ... &  & -0.24 & 0.17 \\
Mg \scriptsize{I} & 6319.5 &  &  & ... &  &  & 0.33 \\
Si \scriptsize{II} & 6347.1 & -0.69 & -0.13 & ... & -0.21 &  & 0.14 \\
Si \scriptsize{II} & 6371.4 & -0.73 &  & ... & -0.36 & -0.33 &  \\
Si \scriptsize{I} & 6721.8 & -0.88 &  & ... & -0.59 &  & -0.01 \\
Si \scriptsize{I} & 6741.6 &  & -0.23 & ... & -0.29 &  & 0.09 \\
Ca \scriptsize{I} & 5260.4 & -0.97 & -0.46 & ... & -0.85 & -0.28 & -0.02 \\ 
Ca \scriptsize{I} & 5349.5 & -0.8 & -0.34 & ... & -0.7 & -0.09 & 0.09 \\
Ca \scriptsize{I} & 6439.1 & -0.88 & -0.49 & ... & -0.56 &  &  \\
Ca \scriptsize{I} & 6471.7 & -0.87 &  & ... & -0.63 &  & 0.1 \\
Ca \scriptsize{I} & 6499.6 & -0.96 & -0.45 & ... & -0.63 & -0.36 & 0.12 \\
Ca \scriptsize{I} & 5261.7 & -0.97 & -0.4 & ... & -0.74 & -0.26 & 0.14 \\
Ca \scriptsize{I} & 6455.6 & -1.01 & -0.48 & ... & -0.7 &  & -0.01 \\
Ca \scriptsize{I} & 6456.9 & -0.46 & -0.1 & ... & -0.29 &  &  \\
Ca \scriptsize{I} & 6493.8 & -0.67 & -0.44 & ... & -0.63 & -0.56 & 0.06 \\
... & ... & ... & ... & ... & ... & ... & ...\\
Ba \scriptsize{II} & 6496.9 & -0.76 &  & ... & -0.67 &  & 0.19 \\
La \scriptsize{II} & 4804.0 &  & 1.03 & ... &  &  & 0.21 \\
La \scriptsize{II} & 5303.5 &  & 0.86 & ... &  &  &  \\
La \scriptsize{II} & 6390.5 &  & 0.88 & ... &  &  & 0.08 \\
Ce \scriptsize{II} & 4914.9 & -0.01 &  & ... &  &  &  \\
Ce \scriptsize{II} & 5117.2 & -0.14 &  & ... &  &  &  \\
Ce \scriptsize{II} & 5187.5 &  &  & ... & -0.62 &  & 0.12 \\
Ce \scriptsize{II} & 5274.2 &  &  & ... &  &  & -0.05 \\
Ce \scriptsize{II} & 5330.6 &  &  & ... &  &  & -0.18 \\
Pr \scriptsize{II} & 5259.6 &  & 0.65 & ... &  &  &  \\
Pr \scriptsize{II} & 5322.7 &  &  & ... &  &  & 0.1 \\
Nd \scriptsize{II} & 4811.3 &  & 1.04 & ... &  &  & 0.29 \\
Nd \scriptsize{II} & 4914.4 &  & 0.75 & ... &  &  &  \\
Nd \scriptsize{II} & 4947.0 &  & 0.59 & ... &  &  & 0.17 \\
Nd \scriptsize{II} & 4989.9 &  & 0.94 & ... &  &  & 0.12 \\
Nd \scriptsize{II} & 5089.8 &  & 0.59 & ... &  &  &  \\
Nd \scriptsize{II} & 5092.8 &  & 0.73 & ... & -0.46 &  &  \\
Nd \scriptsize{II} & 5234.2 &  & 0.68 & ... &  &  &  \\
Nd \scriptsize{II} & 5293.2 &  & 0.74 & ... &  &  &  \\
Nd \scriptsize{II} & 5311.5 &  & 0.78 & ... &  &  &  \\
Nd \scriptsize{II} & 5319.8 & -0.84 &  & ... &  &  &  \\
Nd \scriptsize{II} & 5357.0 &  & 0.83 & ... &  &  &  \\
Nd \scriptsize{II} & 6365.5 &  & 0.65 & ... &  &  &  \\
Eu \scriptsize{II} & 6645.1 &  & -0.05 & ... &  &  &  \\
    \end{tabular}
    \caption{BACCHUS-determined line-by-line abundances (reported as [X/H], assuming solar abundances from \citealt{Grevesse2007}) for our PEPSI spectra.  Final abundances and uncertainties are determined by taking the mean and standard deviation across all lines, respectively.  The full table is available in the published version of this manuscript.}
    \label{tab:bacchus_line_by_line_abunds}
\end{table*}

\begin{figure*}
    \centering
    \includegraphics[width=\linewidth]{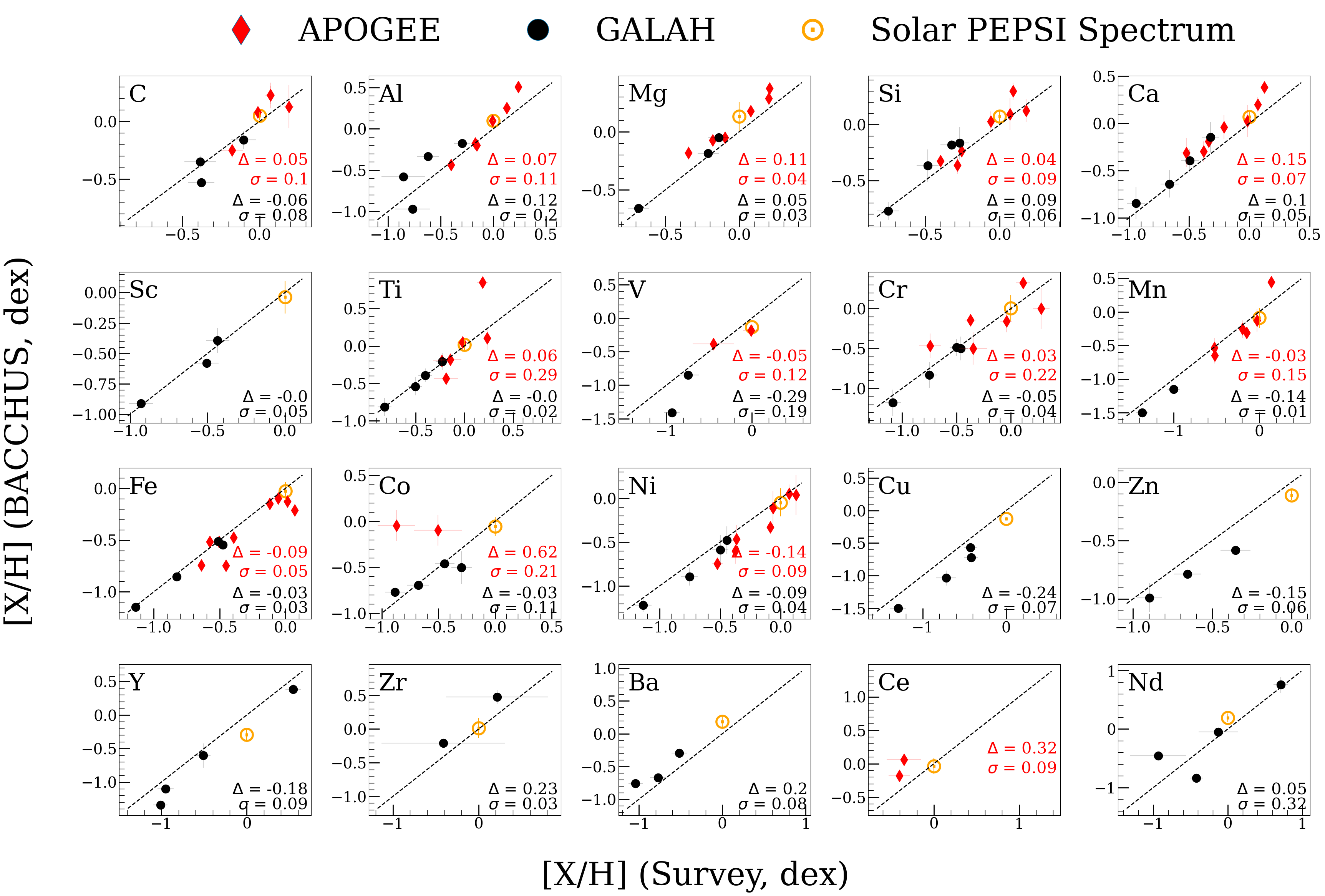}
    \caption{Fitting results for PEPSI using \texttt{BACCHUS} compared to results from APOGEE DR17 \citep{APOGEEDR17} in red and GALAH DR3 \citep{Galahdr3} in black. 
    The bias ($\Delta$) and variance ($\sigma$) are shown in each panel.
    Abundances measured from a PEPSI solar spectrum are also shown for reference to track potential abundance offsets.}
    \label{fig:1}
\end{figure*}

Figure~\ref{fig:1} compares our abundances determined from the PEPSI spectra to those from the published catalogs of APOGEE DR17 \citep{APOGEEDR17} and GALAH DR3 \citep{Galahdr3, Wang2024}  For reference, we also include our abundance determinations for a PEPSI spectrum of the Sun from \citet{Strassmeier18} that we degraded to R=50,000 with iSpec and trimmed to match the wavelength limits of the rest of our spectra.  We observe non-negligible abundance offsets in some elements (Mg, Ca, Fe, Ni, Cu, Zn, Y, Zr, Ba, and Ce) between our PEPSI results and those reported by APOGEE and GALAH.  These offsets are also seen in our analysis of the Solar PEPSI spectrum and can be explained by differences in the spectral data, line selections, model atmospheres, adopted microturbulences, and abundance determination methods between this work and those of APOGEE and GALAH \citep[e.g.,][]{Bedell2014, Ness2015, Xiang2019}.  Upon correcting for these offsets using the solar abundance offsets, we find that [Mg, Ca, Si, and Fe/H], our primary elements of interest for this portion of the analysis, agree with survey abundances within uncertainties (0.03 $\leq \sigma \leq$ 0.09 dex).  Thus, our PEPSI analysis validates the high [$\alpha$/Fe] abundances reported by GALAH and APOGEE for this sample.  The remaining elemental abundances also typically agree with survey results within abundance uncertainties after correcting for offsets with some exceptions (V, Co, and Nd).

%\textcolor{red}{maybe replace with Madi's analysis for Li}
The Li analysis was done independently and used the GALAH Li linelist.
To do so, we used \texttt{Korg}, a fast 1D LTE spectral synthesis code \citep{korg}.
We adopted the log$g$ and \teff\ values from APOGEE DR17 or GALAH DR3 catalogs as inputs to \texttt{Korg} to increase efficiency.
We then fit the rest of the global parameters, namely metallicity ([M/H]), microturbulence ($v$ mic), and rotation broadening ($v$sin$i$).
Finally, we fit for Li while fixing the global parameters.
The linelist is taken to be the same as GALAH DR3 \citep{Galahdr3}, as it includes the Li lines and various iron lines around it.
The Li abundance A(Li) from \texttt{Korg} is calculated using [Li/H]+1.05, that is, with the Solar Li abundance from \citet{Asplund2009}.
The bias and variances for each measurement are shown in the figure legends.
Even though there exist offsets between the surveys, the variance in the abundances is relatively small (typically $<$ 0.1 dex).
Again, these offsets are expected, as differences of up to $\sim$0.1 dex are common between surveys due to variations in instruments, reduction pipelines, wavelength coverage, and fitting methodologies \citep[e.g.,][]{Bedell2014, Ness2015, Xiang2019}.
The best-fit model and data for the Li, Ca, and Mg lines are shown in the middle row in their respective figures (Figure~\ref{fig:3}-\ref{fig:6}).
The data around the H$\alpha$ line is also shown for 18432994+4356589, as emission is seen in the core of the line (Figure~\ref{fig:6}). 

%The disagreement in $v \sin i$ and $v_\mathrm{mic}$ may be due to the absence of macroturbulence in this analysis, or due to slightly varying handling of the instrumental line spread functions.
%We also measured the individual abundances to understand the formation pathways.
%To do so ... \textcolor{red}{Catherine}

%\begin{figure*}
%    \centering
%    \includegraphics[width=\linewidth]{Figures/Abund_comp_catalog.png}
%    \caption{Fitting results for PEPSI using \texttt{Korg} with line lists from GALAH DR3 \citep{Galahdr3} compared to results from APOGEE DR17 \citep{APOGEEDR17} and GALAH DR3 \citep{Galahdr3, Wang2024}. 
%    The bias and variance are shown in the figure legends.
%    The variance for [Fe/H] and [Mg/Fe] agrees within $\sim$ 0.1 dex.}
%    \label{fig:1}
%\end{figure*}

\subsection{Rotation Period Measurements}\label{subsec:period}
For the young \halpha\ candidate stars, we downloaded the light curves using \texttt{lightkurve} \citep{lightkurve2018}.
The rotation periods are determined following the same procedure as in \cite{Lu2025}. 
They are measured using the Lomb-Scargle Periodogram implemented in \texttt{astropy} \citep{astropy:2013, astropy:2018, astropy2022}, with an over-sampling rate around the peak to be 100. 
The raw light curve, periodogram, and the folded light curve on the best detected period are shown in the top row three plots in their respective figures (Figure~\ref{fig:3}-\ref{fig:6}).
For the only star with Kepler data, 18432994+4356589, we perform and show the analysis for two different sectors (Figure~\ref{fig:6}).

\subsection{Initial rejections}\label{subsec:initial_exclude}
For the APOGEE-Kepler/K2 sample, since the binary information is mostly known from multiple RV epochs, the purpose of the spectral follow-up would be to vet for stars with Li measurements in support that the they have not gone through stellar merger or mass transfer events. 
For this purpose, we rejected 7 out of 9 stars without significant Li detection.
However, it is worth pointing out that even solar-age stars can have a range of Li depletion factors \citep[e.g.,][]{Jones1999}, for example, the Sun has a extremely low Li measurement, even though it is only 5 Gyr old.
As a result, some of these stars we rejected may still be young, but we cannot prove this with Li. 
However, a high abundance is a reliable indicator of youth.
Out of the rejected stars, we still analyzed 18432994+4356589 in detail in Section~\ref{subsec:merger} as an potential binary product as it is confirmed to be a \halpha\ star, exhibits fast rotation and \halpha\ emission, but does not have Li detection.

\begin{figure*}
    \centering
    \includegraphics[width=\linewidth]{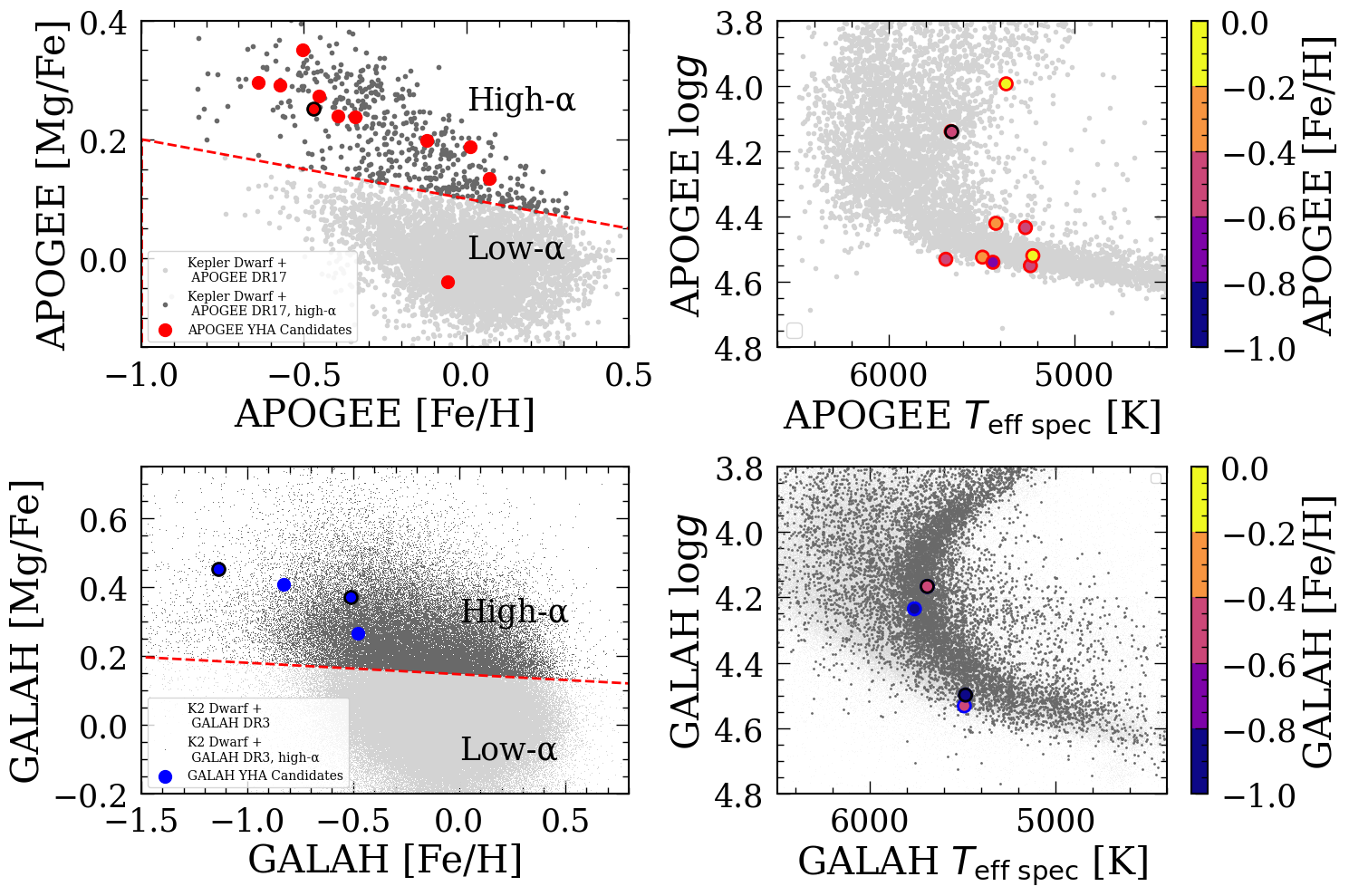}
    \caption{The left column shows the [Mg/Fe]-[Fe/H] plane for the APOGEE sample (top left) and GALAH sample (bottom left). 
    The background gray points show the Kepler dwarf stars in the APOGEE field or the K2 dwarfs in the GALAH field for reference.
    The dark gray points show the \halpha\ stars in each survey for reference.
    The points show the [Mg/Fe] and [Fe/H] values from APOGEE DR17 and GALAH DR3. 
    The red dashed lines --- [Mg/Fe] = $-$0.1[Fe/H]+0.1 for APOGEE, and [Mg/Fe] = $-$0.03[Fe/H]+0.15 for GALAH --- show our separation for the high- and low-$\alpha$ disk, using the background reference.
    The right column shows the Kiel diagram for the APOGEE sample (top right) and GALAH sample (bottom right).
    Again, the background gray points show the reference stars, the dark gray points show the \halpha\ stars with $-0.6$ dex $<$ [Fe/H] $<-0.4$ dex, and the larger points show the values from the public survey data releases for the PEPSI stars, colored by their metallicity values.
    Since we fixed \logg\ and \teff\ while fitting the PEPSI spectra, we do not have a different set of values from our new observations. 
    The points outlined in black show the abundance measurements for the truly young \halpha\ stars that will be described in Section~\ref{subsec:truly young}.
    }
    \label{fig:2}
\end{figure*}

Figure~\ref{fig:2} left column shows our definition of the \halpha\ disk (red dashed lines), abundance measurements from public surveys (red crosses: APOGEE, blue crosses: GALAH), and PEPSI (points).
We defined the line separately for GALAH and APOGEE as their [Mg/Fe]-[Fe/H] plane exhibits slightly different structures.
Using this criterion, we further excluded 19101724+4944353 from our APOGEE sample.
Therefore, we are left with 03413908+2359163 from the APOGEE-Kepler/K2 sample.
The black outlined points show the genuinely young dwarf stars that we will discuss more in detail in Section~\ref{subsec:truly young}.

After the initial exclusion, we only performed detailed analysis, including the binary analysis (Section~\ref{subsec:binary_analy}) and comparison with doppelgänger stars (Section~\ref{subsec:abund_doppel}), for 03413908+2359163, 03435417+1719250, 01084954-0030464, and 18432994+4356589.
These stars are shown as the top 4 entries in Table~\ref{tab:1}.

The right column in Figure~\ref{fig:2} shows the Kiel diagrams.
The points show the stars with follow-up PEPSI observations, and the black outlined points show the truly young dwarf stars, as seen in the [Mg/Fe]-[Fe/H] plots.
We also plotted \halpha\ stars with $-0.6$ dex $<$[Fe/H]$<-0.4$ dex in dark gray points for comparison.
The fact that some of the points do not hug the left edge of the distribution of the dark gray points suggest they are likely younger than the full \halpha\ population.
However, for the main-sequence stars, the isochrone tracks overlap heavily, and it is known that isochrone ages are not accurate for these stars, and worse, often overestimate their ages by more than 50\% \citep{Byrom2024}.
Moreover, 03413908+2359163, the one truly young dwarf star in the APOGEE-K2 sample outlined in black in the top row has a high reddening value, E(\bprp) $\sim$ 0.5, estimated using the 3D dustmap \texttt{bayestar2019} \citep{Green2019}, implemented in \texttt{dustmap} \citep{Green2018, Green20182}.
This high reddening value could also affect the estimations of \teff, \logg, and age for this star.

\subsection{Binary Analysis}\label{subsec:binary_analy}
We use RVs and Gaia RUWE to place constraints on possible companions to each candidate, using an updated version of the Multi Observational Limits on Unseen Stellar Companions \citep[\texttt{MOLUSC};][]{wood2021} tool.
\footnote{\url{https://github.com/stephtdouglas/MOLUSC}, \url{https://github.com/stephtdouglas/lu_highalpha}} 
\texttt{MOLUSC} uses rejection sampling to infer the posterior probability distribution on binary companion parameters: orbital period $P_{orb}$, eccentricity $e$, inclination $i$, mass ratio $q$, argument of periapsis $\omega$, pericenter phase $\phi$, and system RV $v_{sys}$.
The prior is flat for $0\le q\le1$, $0\le\phi\le\pi$, $0\le\omega\le2\pi$, and $0\le\cos i\le1$. 
We use a log-flat prior on $P_{orb}$. For $e$, we use a normal distribution for $P_{orb}<1000$~d and a uniform distribution for $P_{orb}\geq1000$~d. The normal distribution has the same parameterization as in \citet{wood2021}:
\begin{equation}
    \mu_e = 0.148 \log_{10}(P_{orb}) + 0.001
\end{equation}
\begin{equation}
    \sigma_e = 0.042 \log_{10}(P_{orb}) + 0.128
\end{equation}

We start with 50~million candidate companions sampled randomly from the prior. 
We calculate the predicted RV semiamplitude $K$ for each simulated companion, and numerically solve the Kepler Equation at each observing time to produce a predicted RV. 
We place an RV floor of 100~m s$^{-1}$ for APOGEE and 50~m s$^{-1}$ for PEPSI; any candidate with predicted variability below this limit is rejected. 
Based on predicted contrast, we also reject spectroscopic binaries with line-splitting (SB2s) that would appear resolved in the observed spectra. 
We then compare the predicted RVs to measured RVs using a $\chi^2$ test, with the $\chi^2$ cumulative distribution function probability taken as the rejection probability of the system. 
For systems with RVs from both PEPSI and APOGEE, we calculate limits using each dataset separately, as the instrumental offset could be mistaken for astrophysical variability in the RV analysis. 
The RV measurements from APOGEE are taken from \cite{Saydjari2025}, which agree with the APOGEE DR17 \texttt{Doppler} outputs \citep{doppler} for our stars.
%We do include it for completeness. \citet{wood2021} derive a sensitivity map of RUWE in terms of Gaia G magnitude contrast ($7\le\Delta G \le0$) and separation ($2\times10^1\le separation (mas)\le2.1\times10^3$. This sensitivity map is used to derive the rejection probability for each candidate companion.
%Candidate companions outside the limit of the RUWE sensitivity map are never rejected.
We then calculate the confidence limit on the presence of a candidate companion by binning the samples in parameter space and calculating the percentage of samples that survived rejection sampling. 
Given the large distance to these stars, the binary signal will not have a significant impact on astrometry, so Gaia RUWE does not produce very significant constraints.

We have done this analysis for the two GALAH-K2 stars with confirmed Li measurements (03435417+1719250 and 01084954-0030464), the one APOGEE-K2 star with Li measurement (03413908+2359163), and the one APOGEE-Kepler star with interesting features (18432994+4356589).
The RVs and binary analysis results are shown in the bottom right panel in their respective figures (Figure~\ref{fig:3}-\ref{fig:6}).
These figures show the 99$^{\rm th}$-percentile confidence limit --- a candidate companion with a mass ratio above this line is rejected with 95\% confidence.
The analysis for each plot will be discussed in Section~\ref{sec:result}.

\subsection{Abundance Comparison with Doppelgänger Stars}\label{subsec:abund_doppel}
To understand whether there are unique patterns in the detailed chemical abundances, including planet engulfment signatures, of the likely young \halpha\ stars, we compare the abundances of these stars with their doppelgängers.
We do so by selecting stars that agree with the target star's \logg, \teff, [Fe/H], and [Mg/Fe], within uncertainty. 
We followed the same procedure as the previous paper \citep{Lu2025}, but enforced an SNR condition so that the differences between the targeted star and the doppelgängers' SNR values are less than 50.
The SNR for GALAH ID 160110002101185, 2MASS J03435417+1719250, is 126; The SNR for GALAH ID 161006004901279, 2MASS J01084954-0030464, is 63.6.
V is missing for GALAH ID 161006004901279, and Al is missing for GALAH ID 160110002101185 because there are no reported GALAH measurements.
The SNR for 2MASS J18432994+4356589 is 74.6; The SNR for 2MASS J03413908+2359163 is 116.6.
We did not use the abundances from PEPSI to perform this test as there are known differences between the surveys, and since we do not have PEPSI observations for the doppelgänger stars, we cannot perform a consistent comparison using abundances from PEPSI.
The results are shown in the bottom left panel in each of their corresponding figures (Figure~\ref{fig:3}-\ref{fig:6}).

\section{Results} \label{sec:result}
Combining abundances (Li, Mg, and Fe), rotation periods, and RV measurements, we identify three genuinely young \halpha\ rotators (see Section~\ref{subsec:truly young}), and one possible merger product (see Section~\ref{subsec:merger}). 
We will now take a look at the combined information for these four stars individually.
These results are also summarized in Table~\ref{tab:2}.

\begin{table*}[]
    \centering
    \begin{tabular}{c|cccccc}
         2MASS ID& \prot\ [Days] & Age [Gyr] & A(Li) & Planet Engulfment? &  Stellar Companion? & $s$-process Enhancements? \\
         \hline
          03413908+2359163 & 32.21 & 5.80$^{+0.13}_{-0.22}$ & 1.39 & N & N & N/A \\   
          03435417+1719250 & 16.52 & 2.36$^{+0.05}_{-0.11}$ & 1.42 & N & N & [Ce/Fe]=1.00 dex \\
          01084954-0030464 & 20.98& 3.84$^{+0.27}_{-0.19}$ & 2.13 & N & N & N
          
    \end{tabular}
    \caption{2MASS ID, rotation periods, gyrochronology age inferred with \texttt{GPgyro}, A(Li) measured using [Li/H]+1.05, with [Li/H] measured from \texttt{Korg}, whether the star exhibit planet engulfment signature from comparing with doppelgänger stars, whether the RV measurements show signature of a close-by stellar companion, and whether it shows enhancements in $s$-process elements according to PEPSI spectra analysis with \texttt{BACCHUS} (to vet for asymptotic giant branch (AGB) mass transfer).
    We only include values of $s$-process element abundances from PEPSI if [X/Fe] $>$ 0.5 dex.
    We are not able to measure any $s$-process elements for 03413908+2359163 due to the low SNR for the PEPSI spectra.}
    \label{tab:2}
\end{table*}

\subsection{Three Genuinely Young \halpha\ Dwarfs}\label{subsec:truly young}
In this section, we will look at the three newly identified genuinely young \halpha\ dwarf stars (Section~\ref{subsubsec:young1} to Section~\ref{subsubsec:young3}) and the one possible merger (Section~\ref{subsec:merger}) identified in Section~\ref{subsec:initial_exclude} individually.
We will then discuss the age and chemical signatures the genuinely young \halpha\ dwarf stars have in common in Section~\ref{subsec:abund_comp}.
Finally, we will speculate the possible formation pathways of genuinely young \halpha\ stars in Section~\ref{subsec:formation}.

\subsubsection{2MASS J03413908+2359163}\label{subsubsec:young1}

\begin{figure*}
    \centering
    \includegraphics[width=\linewidth]{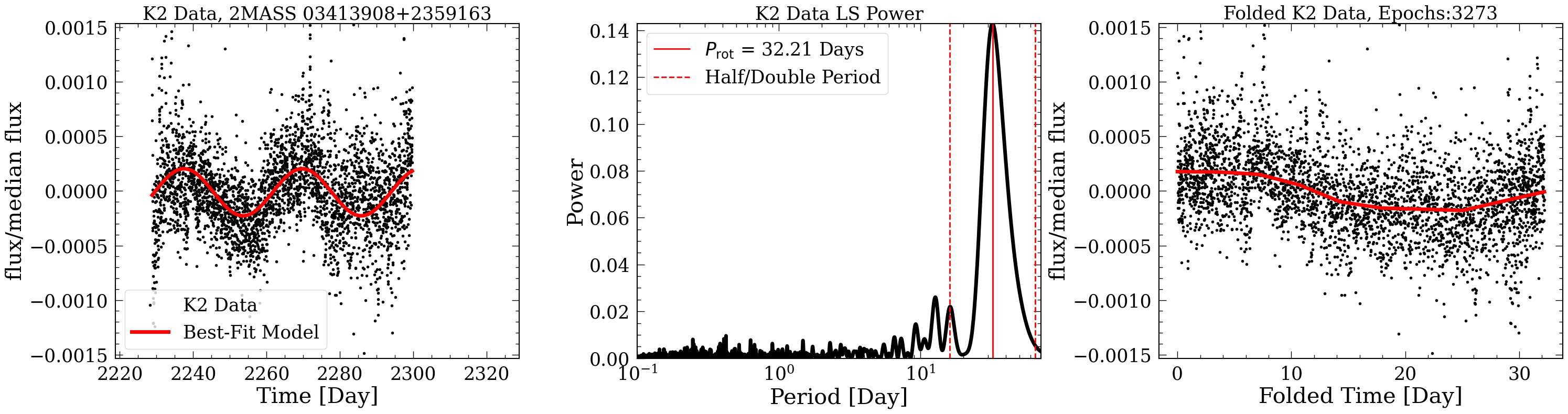}
    \includegraphics[width=0.33\linewidth]{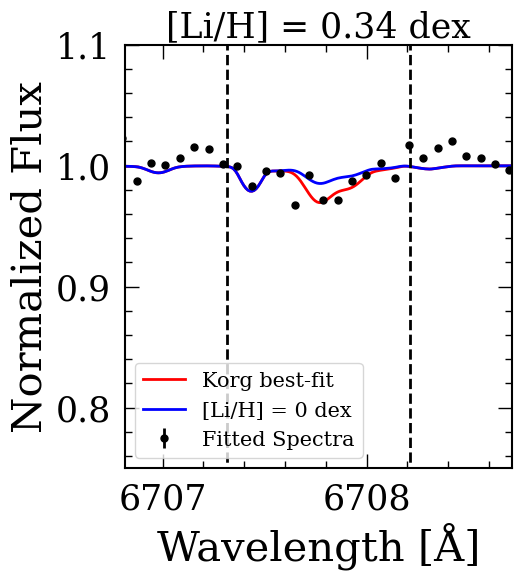}
    \includegraphics[width=0.66\linewidth]{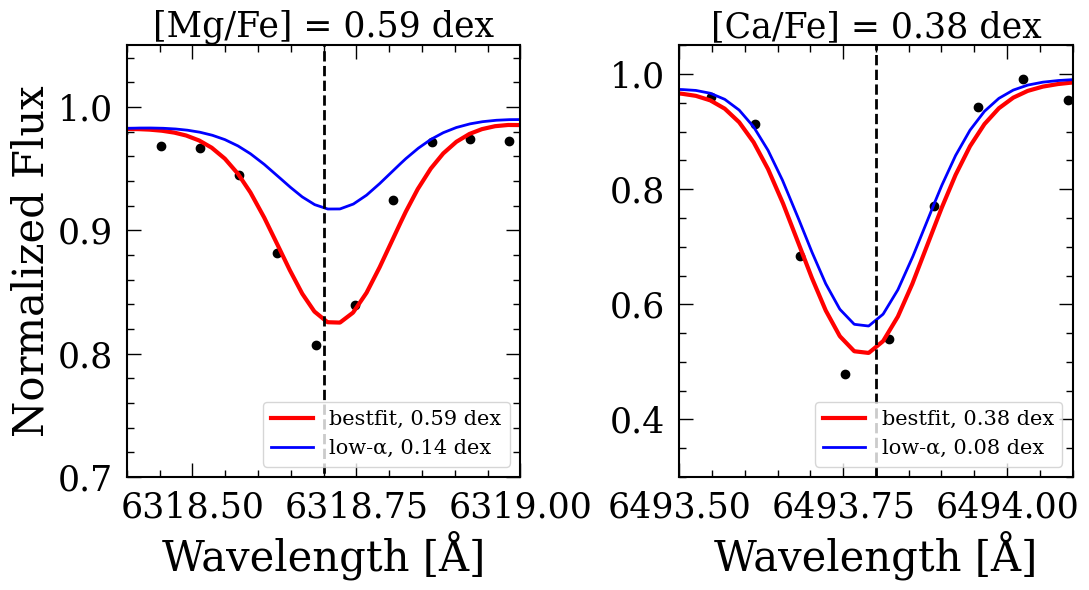}
    \includegraphics[width=0.49\linewidth]{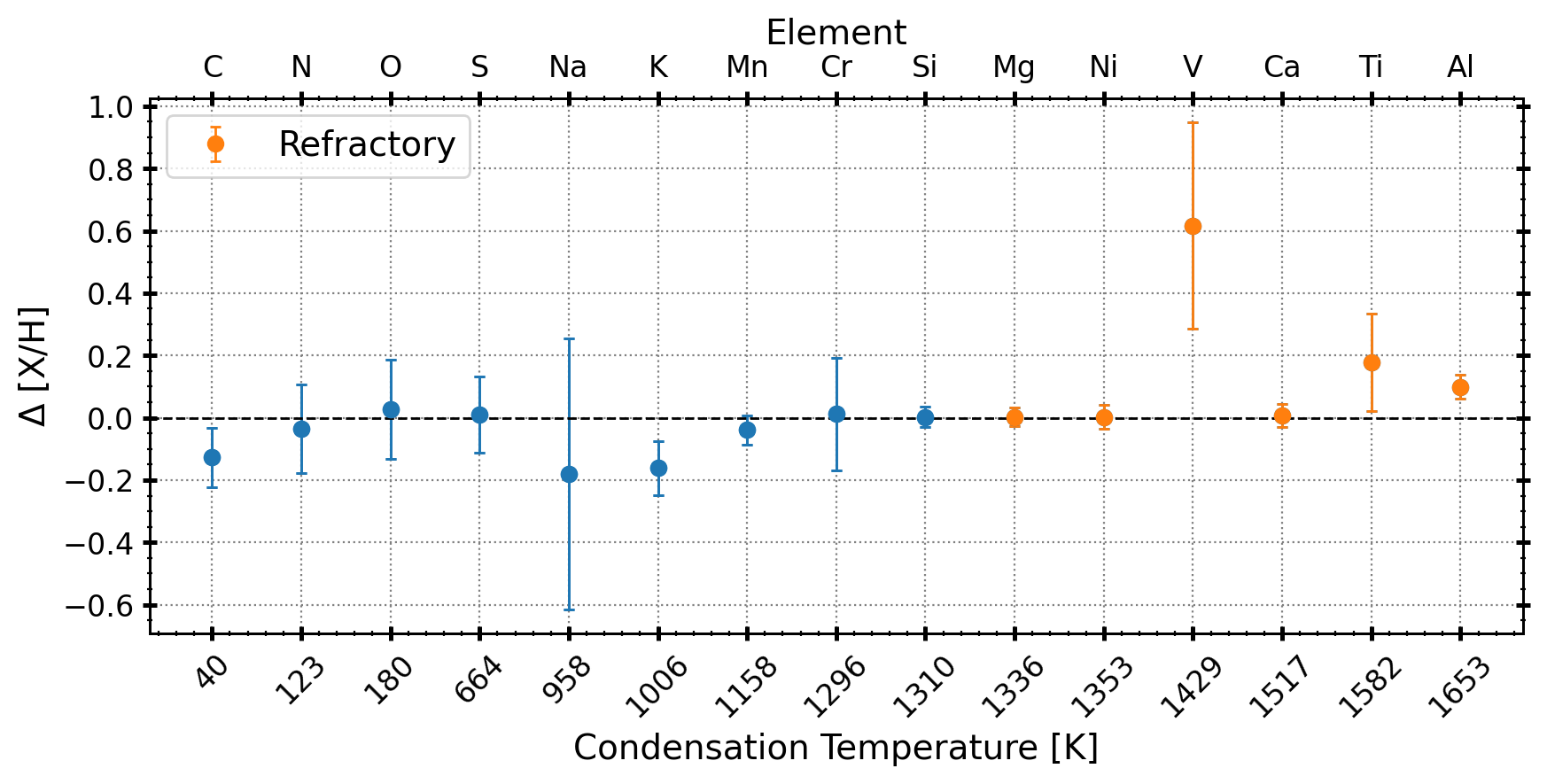} 
    \includegraphics[width=0.245\linewidth]{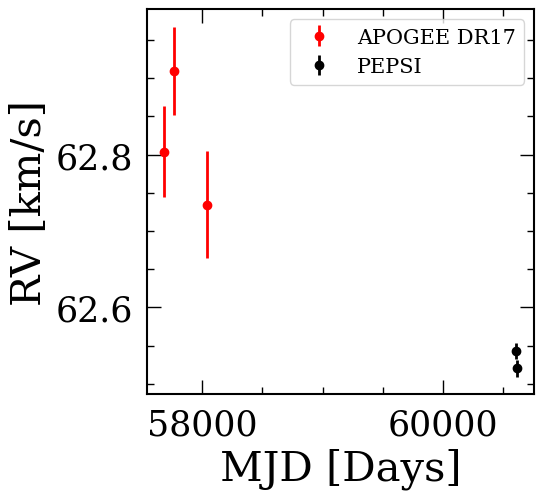}
    \includegraphics[width=0.25\linewidth]{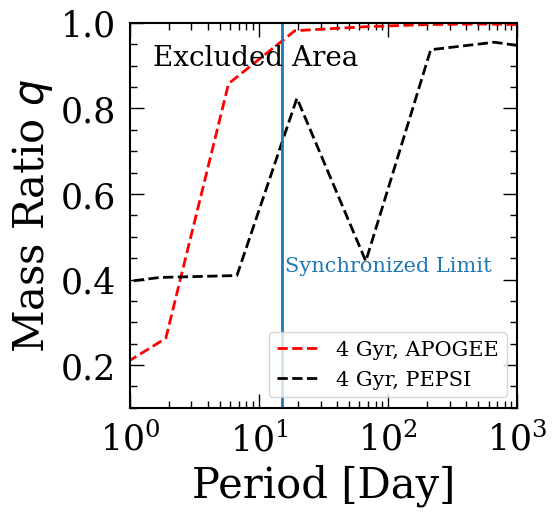}
    \caption{Detail analysis for 03413908+2359163. Top: K2 light curve with the best-fit model (left), the Lomb-Scargle Periodogram where the red line shows the period detected and the red dotted lines show the half and double periods (middle), and the folded light curve onto the detected period (right). 
    Middle left: \texttt{Korg} best-fit model (red lines) around the Li line.
    The vertical black dashed lines show the window for the Li line.
    The black points show the PEPSI spectra for this targeted star, and the blue line shows the synthetic spectrum of a star with the same stellar parameters but with [Li/H] = 0 dex.
    Middle right: \texttt{BACCHUS} best-fit model (red) to the data for the Mg 6318.7 line and the Ca 6493.8 line. 
    The vertical dashed line shows the centers of the lines.
    The blue lines show the synthetic spectra of a low-$\alpha$ star with the same stellar parameters generated from \texttt{BACCHUS} for comparison. 
    Bottom left: Abundance comparisons with doppelgänger stars (detail see Section~\ref{subsec:abund_doppel}).
    The orange points show refractory elements used to vet planet engulfment events.
    Bottom right: RV measurements and binary exclusion tests for this star.
    In the binary exclusion test plot showing mass ratio vs orbital period, the blue vertical line shows the synchronized binary limit at 15 days, binary stars with an orbital period less than 15 days should be synchronized.
    The lines show 95\% confidence level for binary exclusion assuming a typical age of a \halpha\ star of 10 Gyr, analyzed with PEPSI RVs (black) and APOGEE RVs (red) separately. 
    The dashed lines show those for an intermediate age \halpha\ star of 4 Gyr.}
    \label{fig:3}
\end{figure*}

This star has a rotation period measurement of 32.21 days, with no other strong peaks detected around the main peak in the periodogram.
With its Gaia \bprp\ color of 1.28, and period measurement, \texttt{GPgyro} \citep{Lu2024} inferred an age of 5.8$^{+0.2}_{-0.2}$ Gyr.
With additional metallicity information, \texttt{STAREVOL} inferred an age of 5.1$^{+0.8}_{-0.1}$ Gyr, agreeing with that from \texttt{GPgyro}.
This is an intermediate age \halpha\ star since typical \halpha\ stars have ages between 7-13 Gyr \citep[e.g.,][]{Xiang2022, Pinsonneault2025}.

This star was observed in APOGEE, with three epochs of RVs available. 
We performed two additional follow-ups with PEPSI.
The binary analysis suggests we are able to exclude a mass ratio of $\sim$0.4 to the synchronized limit with APOGEE RVs.
This is about 0.34 \Mdot, using the output mass of 0.86 \Mdot\ for the primary star from \texttt{STAREVOL}.
However, if the star has a stellar companion with a mass ratio of 0.4, is it likely already synchronized by the typical age of a \halpha\ star of 8 Gyr \citep[assuming the equilibrium tide prescription using Equation 4.6 of][]{Zahn1977}.
As a result, if an actual companion exists, its mass ratio is likely less than 0.1.
The binary analysis suggests this star likely does not have a stellar mass donor, and have not been spun up by close-by stellar companions.
Future follow-up should focus on collecting more RV measurements to place further exclude sub-stellar companions, if one exists.
%\textcolor{red}{can we place some upper limit on how much spin up it has had}

Abundance analysis shows significant Li abundance in the star's atmosphere (see middle left panel in Figure~\ref{fig:3}), suggesting it has not gone through stellar mergers.
Although one may argue planet engulfment and pollution from nearby AGB or novae can also lead to enhancement in surface Li for unevolved stars \citep[e.g.,][]{Koch2011, Li2018, Matsuno2025}, these unevolved stars discovered previously are typically metal-poor, with [Fe/H] $<$ -1 dex, and have enhancements in $s$-process elements, Na, or N, which are not seen in any of our stars. 
Doppelgänger analysis (bottom left panel) shows no strong trend with refractory elements or condensation temperature, suggesting no obvious signature of planet engulfment.

\subsubsection{2MASS J03435417+1719250}\label{subsubsec:young2}
\begin{figure*}
    \centering
    \includegraphics[width=\linewidth]{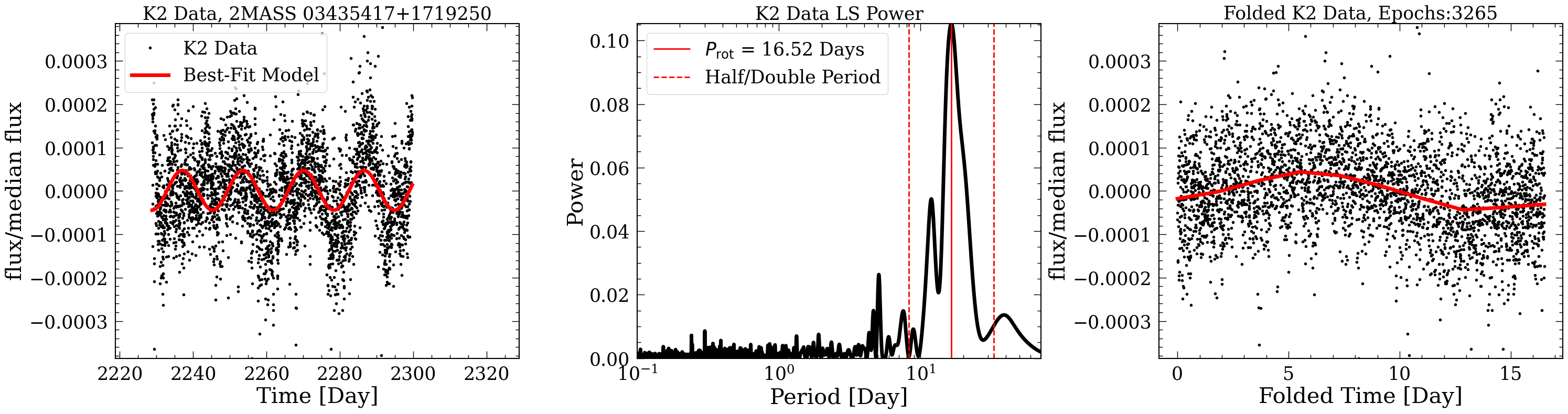}
    \includegraphics[width=0.33\linewidth]{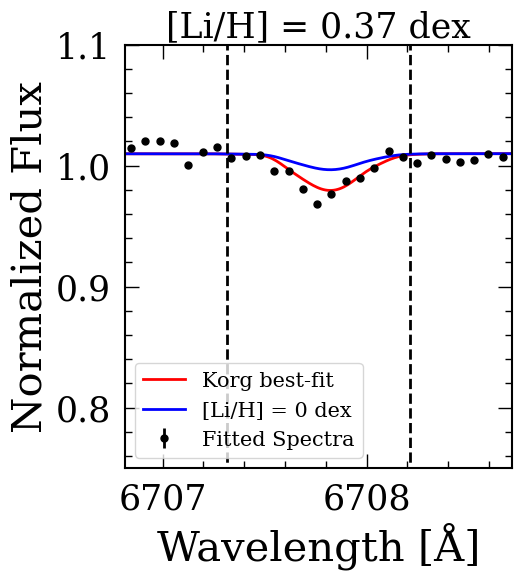}
    \includegraphics[width=0.66\linewidth]{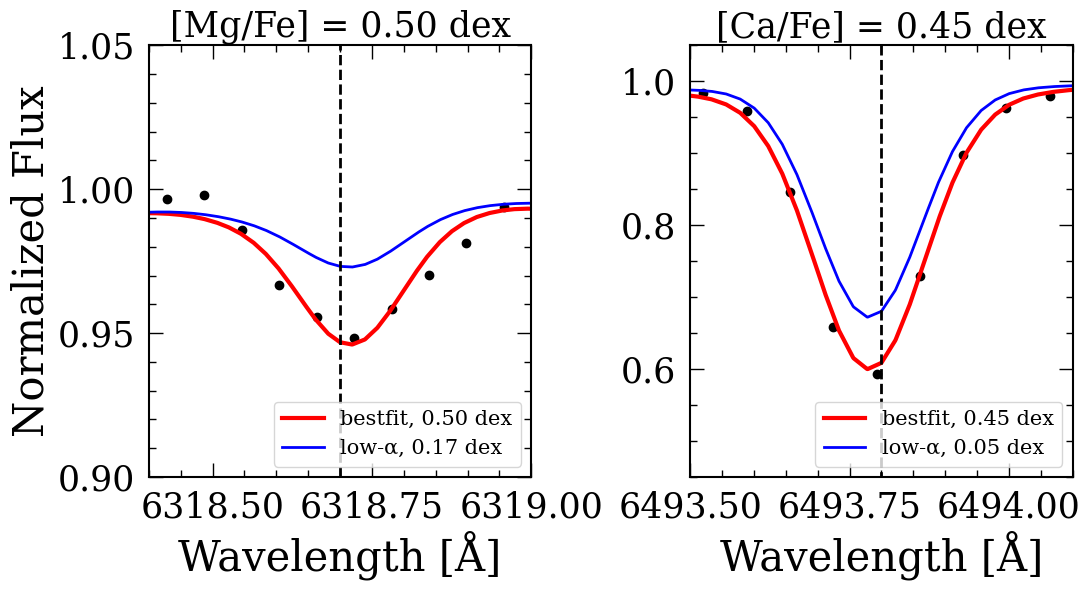}
    \includegraphics[width=0.47\linewidth]{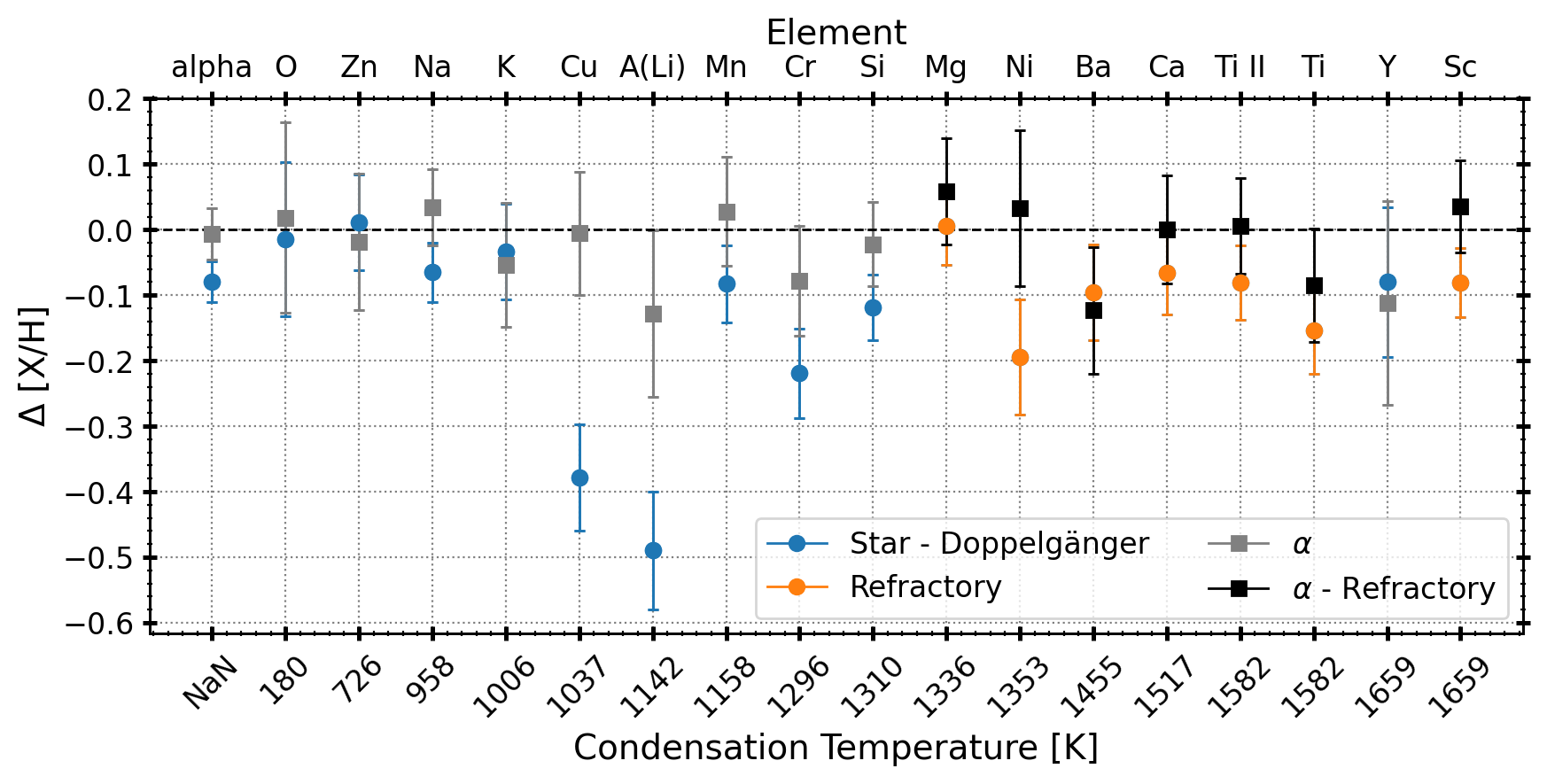} 
    \includegraphics[width=0.255\linewidth]{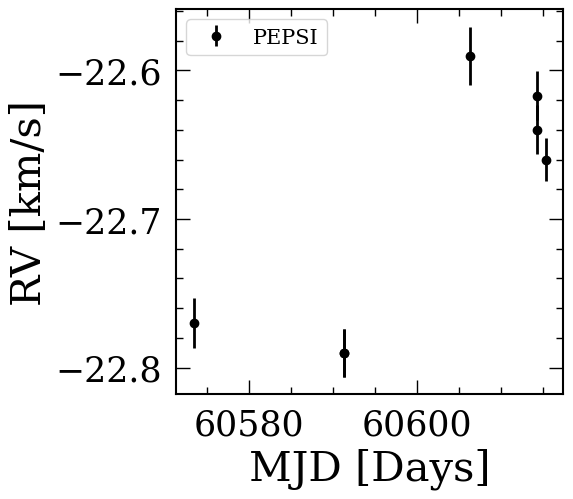} 
    \includegraphics[width=0.25\linewidth]{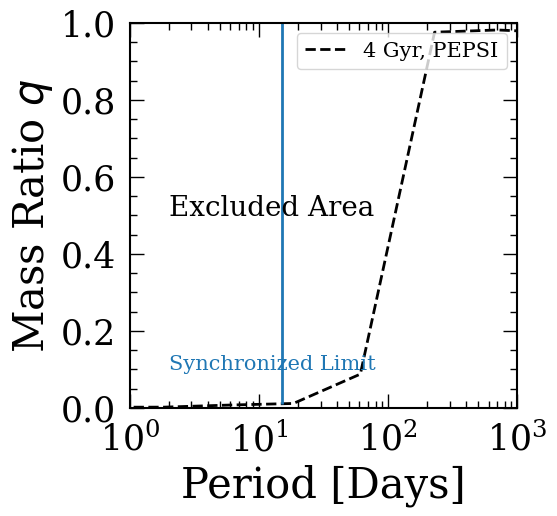}
    \caption{Similar as Figure~\ref{fig:3} but for 03435417+1719250.
    The differences here is for the comparison with doppelgänger stars in the bottom left panel, we compared both with doppelgängers selecting on agreeing [Mg/Fe] (circle points) or agreeing \alphafe\ (square points) as this star likely exhibit an abnormally high [Mg/Fe] values.
    This is a GALAH-K2 candidate star.}
    \label{fig:4}
\end{figure*}
This star has a clear rotation signal at 16.52 days, with no strong peaks at the half or double period.
With its Gaia \bprp\ color of 0.82, and period measurement, \texttt{GPgyro} \citep{Lu2024} inferred an age of 2.35$^{+0.07}_{-0.09}$ Gyr.
\texttt{STAREVOL} is not able to provide precise age information for this star, after combining isochrone tracks with its rotation period.

With 6 RV measurements from PEPSI, we are better able to exclude binary companions down to a mass ratio of 0.04 at the synchronized period.
This mass ratio excludes a close companion over 0.017 \Mdot (about 17 $M_{\rm J}$), using the output mass of 0.83 \Mdot\ for the primary star from \texttt{STAREVOL}.
As a result, no close-in companion may be able to affect its rotation period, since a hot Jupiter does not strongly affect its host star's period evolution tidally beyond 0.05 AU \citep[e.g.,][]{Benbakoura2019}.

This star also has a detectable Li abundance as shown in the middle left panel in Figure~\ref{fig:4}.
Interestingly, by selecting on [Mg/H] as an indication for $\alpha$ element, almost all the other $\alpha$ elements for this star are lower compared to its doppelgängers.
If this star has a slightly elevated Mg, it could account for some of the discrepancy between the element abundances. 
Indeed, if we select doppelgänger stars based on the overall $\alpha$ elements, the differences in element abundances disappeared or decreased significantly, including Li and Cu, which exhibit the most obvious discrepancy when selecting doppelgängers via [Mg/H]. 
This is the only star in our genuinely young GALAH--K2 sample that does not show elevated Li abundance.
This star also has the lowest metallicity compared to the rest. 

\subsubsection{2MASS J01084954-0030464}\label{subsubsec:young3}
\begin{figure*}
    \centering
    \includegraphics[width=\linewidth]{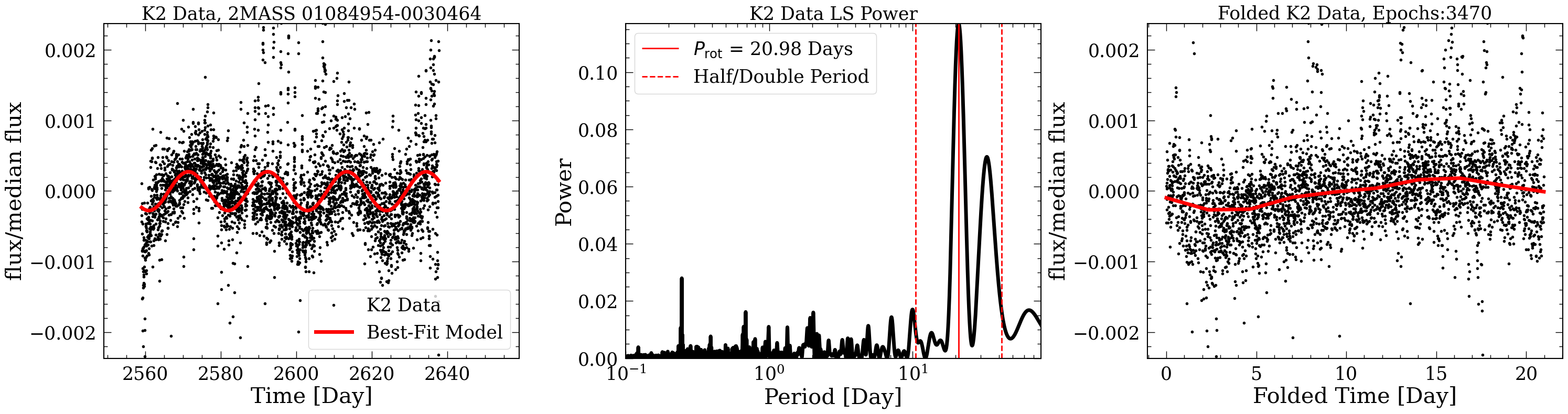}
    \includegraphics[width=0.33\linewidth]{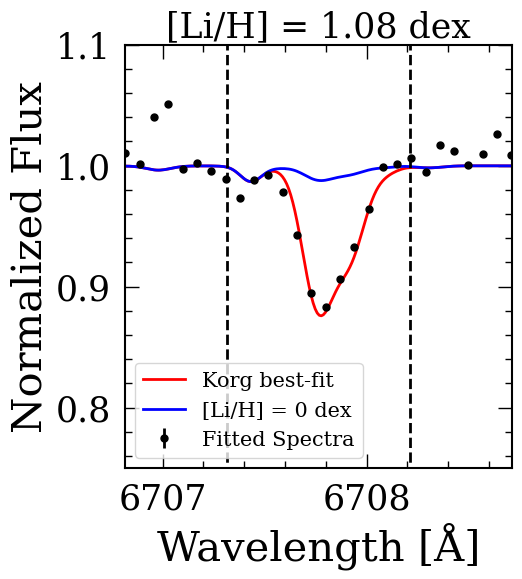}
    \includegraphics[width=0.66\linewidth]{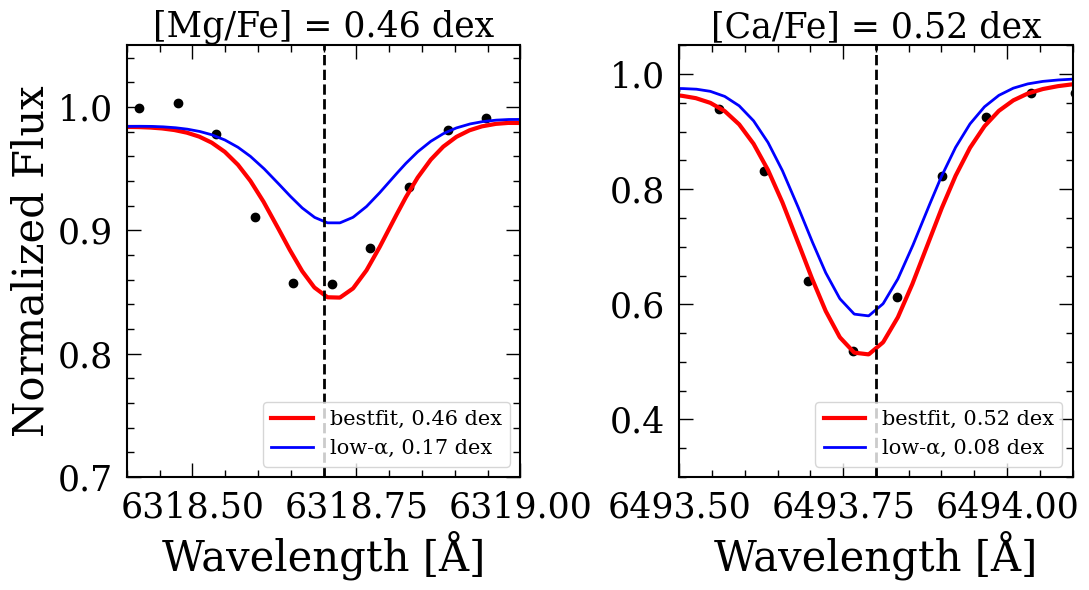}
    \includegraphics[width=0.48\linewidth]{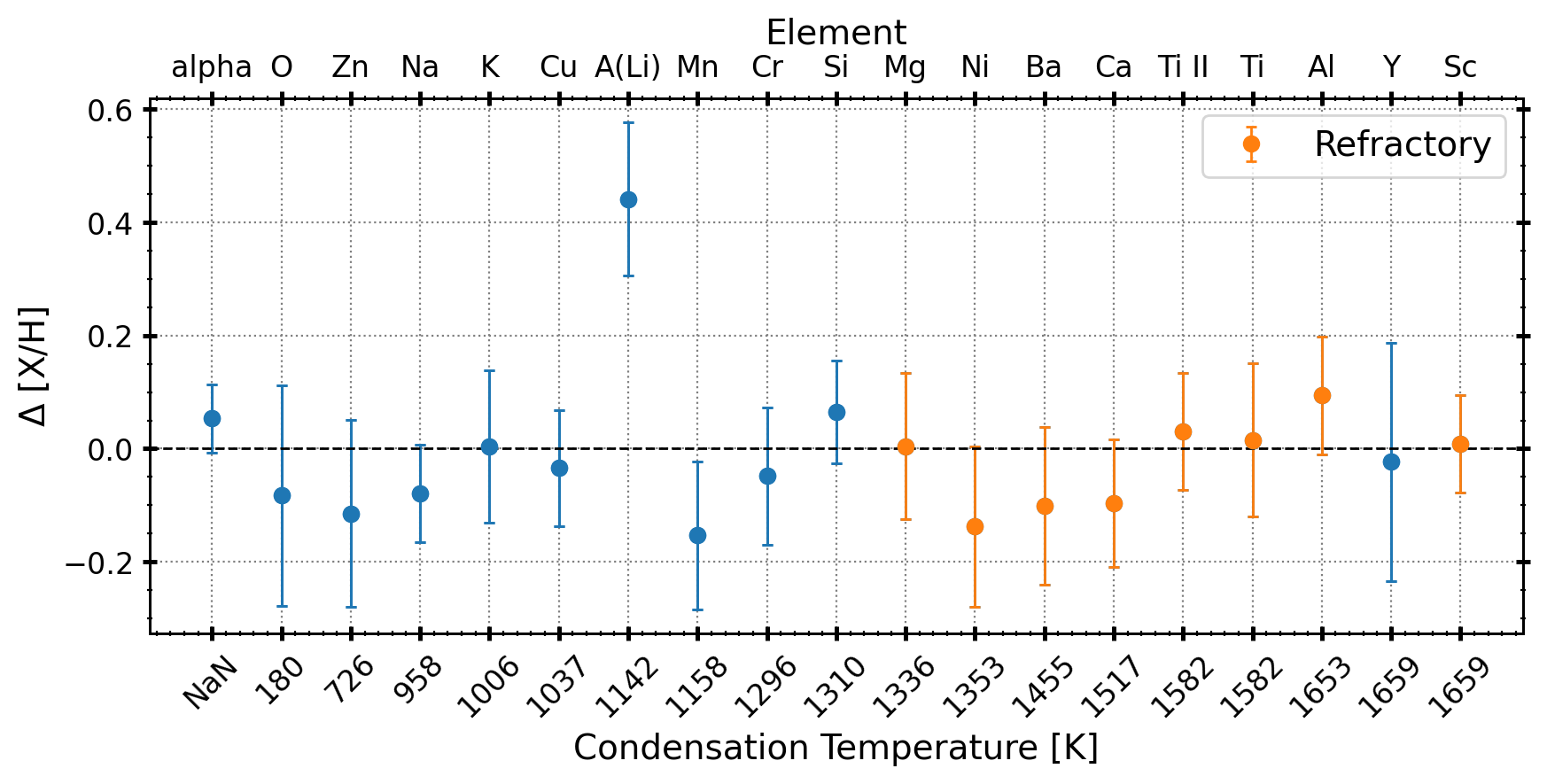} 
    \includegraphics[width=0.26\linewidth]{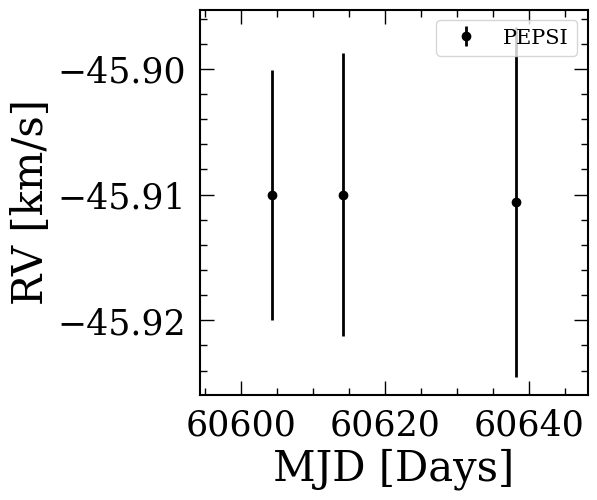} 
    \includegraphics[width=0.24\linewidth]{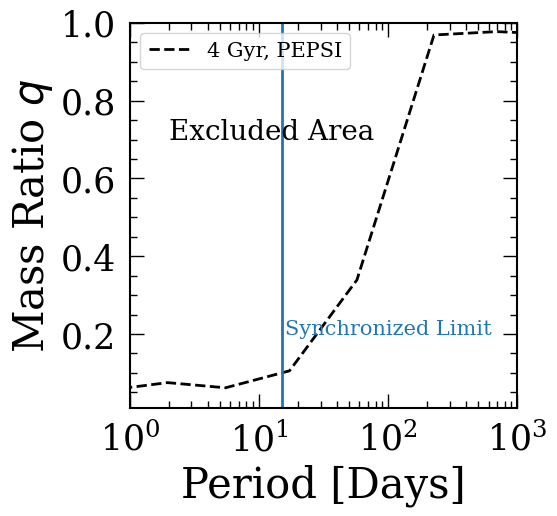}
    \caption{Similar as Figure~\ref{fig:3} but for 01084954-0030464.
    Here we use the Ca 6439.1 line.
    This is a GALAH-K2 candidate star.}
    \label{fig:5}
\end{figure*}
Period detection for this star has the strongest peak at 20.98 days (see top middle panel from Figure~\ref{fig:5}), which yields an age of 3.9$^{+0.3}_{-0.2}$ Gyr with \texttt{GPgyro}.
Again, with its unique combination of stellar parameters, \texttt{STAREVOL} is not able to provide precise age information for this star.
However, a possible peak exists around double the period at 40 days.
If the second highest peak is the true period ($\sim$35 days), this star would be 11.9$^{+1.8}_{-1.4}$ Gyr according to \texttt{GPgyro}, which matches with that of a normal \halpha\ star.

This star shows an elevated Li abundance compared to its doppelgänger stars (bottom left panel in Figure~\ref{fig:5}), and binary analysis suggests no high mass companions (mass ratio $>$ 0.5) exist within an orbital period of 100 days. 
No strong trend differences in abundances exist besides Li.
However, a possible positive slope exists with condensation temperature for refractory elements as shown in the bottom left panel of Figure~\ref{fig:5}.
This trend, if real, could suggest a recent planet engulfment event \citep[e.g.,][]{Ramirez2009, Teske2016, Melendez2017, Spina2021}.
Moreover, it has a relatively high [Ce/Fe] measurement of 1.0 dex, suggesting possible AGB mass transfer.
However, this is unlikely as we do not see any nearby stellar companion.

\subsection{Patterns in Element Abundances for Genuinely Young \halpha\ Dwarfs}\label{subsec:abund_comp}
Combining the other genuinely young \halpha\ dwarfs from \cite{Lu2025}, we are able to compare their abundances with the doppelgänger stars for these four stars altogether, in order to understand whether they exhibit unique chemical patterns.
Note, the comparisons are done in their respective surveys (i.e., if the targeted star is in GALAH, we select doppelgänger stars only from GALAH and compare abundances from the GALAH survey) to avoid biases across surveys.

\begin{figure*}
    \centering
    \includegraphics[width=0.9\linewidth]{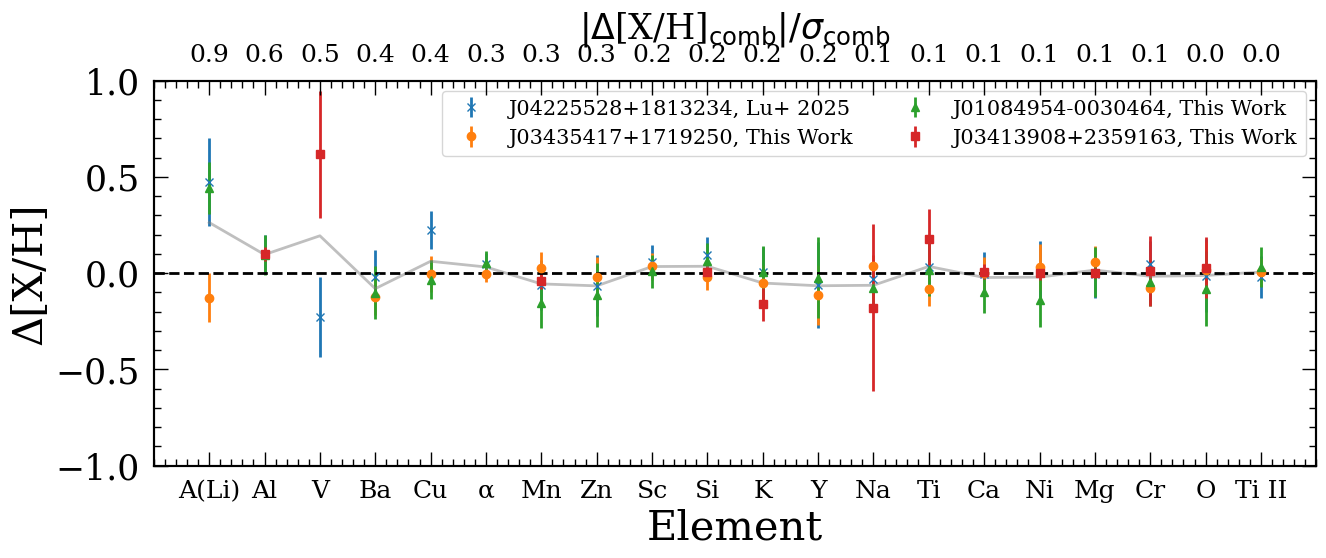}
    \includegraphics[width=0.9\linewidth]{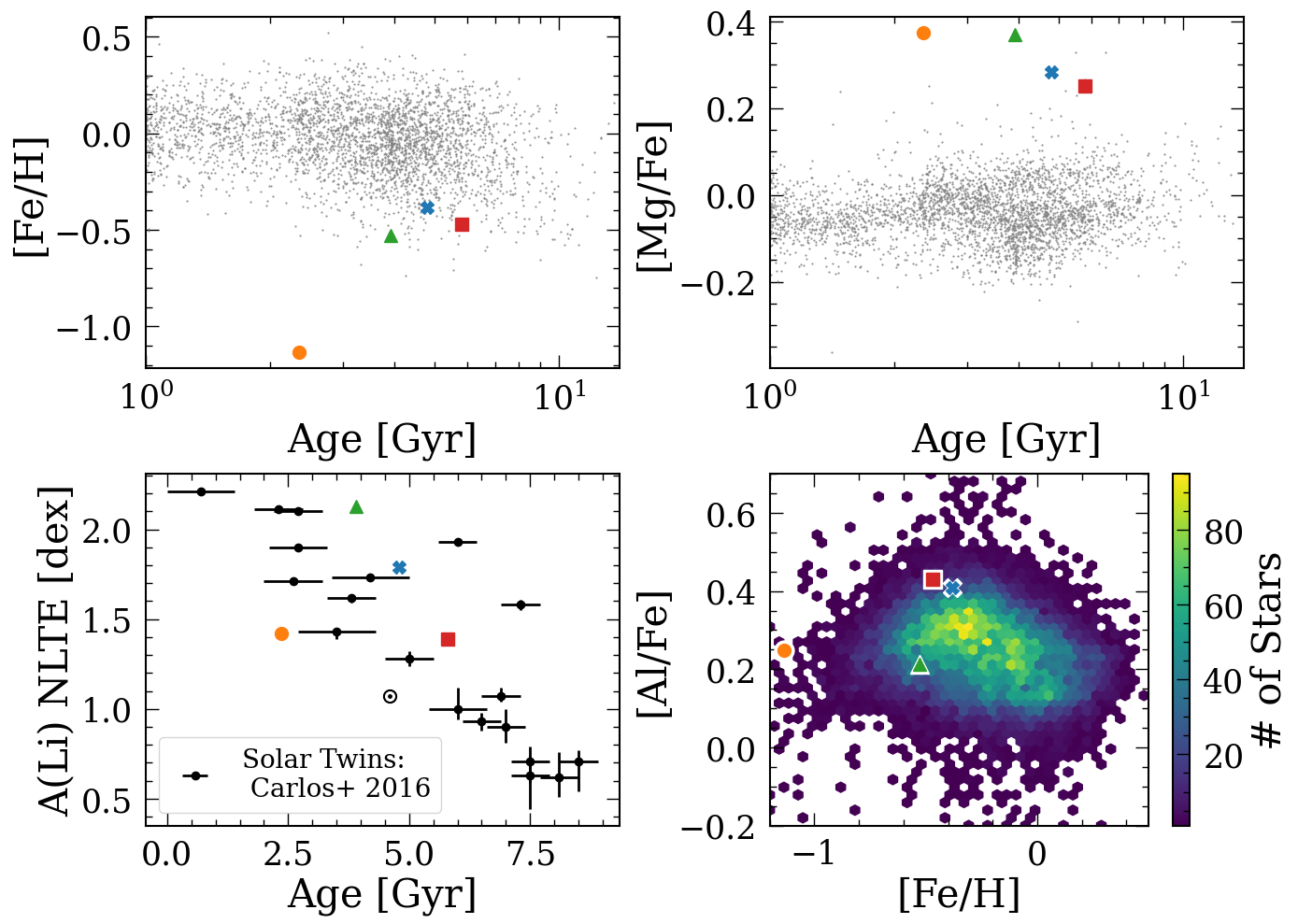}
    \caption{Top Row: combined element abundance difference between each genuinely young \halpha\ dwarf star and its doppelgängers ordered by decreasing absolute average differences, normalized by the average uncertainty (shown as the top axis). 
    The gray line shows the average differences for each element.
    Interestingly enough, three out of four stars are $\sim$0.1 dex elevated in [Al/H] compared to their doppelgängers, except 03435417+1719250.
    Two of the three stars in GALAH have Li measurements about 0.5 dex higher than their doppelgängers.
    Middle Row: [Fe/H]-age and [Mg/Fe]-age relation for all rotating dwarf stars in APOGEE DR17 (gray) with ages determined from \cite{Lu2024}, and the four genuinely young \halpha\ dwarf star (colored symbols, as indicated in the caption in the top plot). 
    Bottom left: Li abundance as a function of age for solar twins taken from \cite{Carlos2016} shown in the black points, and our sample of 4 stars are shown in color points, and the Sun (``$\odot$'' symbol) is also shown.
    Our 4 stars except 03435417+1719250 follow the expected Li-age trend, suggesting the Li contend of these stars has likely not been altered by binary mass transfer.
    This may be due to the fact that 03435417+1719250 has a significantly lower measured metallicity.
    Bottom right: [Al/Fe]-[Fe/H] relation for all the \halpha\ dwarf stars in APOGEE DR17 (\logg\ $>$ 4, \teff\ between 4500 and 6500 K, with no flags) in the background histogram, and the colored points with white outlines are the three out of four genuinely young \halpha\ dwarf star.
    Al measurement is missing for 03435417+1719250 in GALAH DR3, and as a result, we adopted the value calculated from BACCHUS.}
    \label{fig:6}
\end{figure*}

Figure~\ref{fig:6} top plot shows the combined element comparisons for these four stars with respect to the doppelgängers.
For 03435417+1719250, we used the doppelgänger stars selected with $\alpha$ instead of [Mg/H] as this star likely has elevated [Mg/H] values (see Section~\ref{subsec:truly young}). 
The elements are ordered in decreasing average difference after accounting for the combined uncertainty.
To do so, we first calculated the absolute value of the average difference ($|\Delta$[X/H]$_{\rm comb}|$), we then divided it by the average uncertainty ($\sigma_{\rm comb})$, calculated by adding the uncertainties for individual stars in quadrature. 
The elements are then ordered by decreasing in $|\Delta$[X/H]$_{\rm comb}|/\sigma_{\rm comb}$, as shown as the top axis.
The gray line shows the average difference, $\Delta$[X/H]$_{\rm comb}$, for each element.

The difference in most $\alpha$ elements (e.g., Ti, Ca, Ni, Mg, O) is the smallest among all stars, which is a sanity check as we selected stars that have [Mg/H] or [$\alpha$/H] agreeing within uncertainty to be the doppelgänger stars.
The most interesting elements are Li and Al, where out of the three GALAH--K2 stars with Li measurements from \cite{Wang2024}, two stars have similar elevated Li measurements, about 0.5 dex, above their doppelgängers.
All three out of four stars with Al measurements also exhibit a similar amount of slightly elevated [Al/H] values compared to their doppelgängers. 
Besides Li and Al, some other abundances, such as V, Ba, and Cu, also show some variations compared to the doppelgängers.
However, it is worth pointing out that V is difficult to measure, and it is often unreliable in GALAH.
In the future, we plan on proposing for PEPSI time to observe the doppelgänger stars for a more in depth comparison. 

The middle row of Figure~\ref{fig:6} shows the [Fe/H]-age and [Mg/Fe]-age relation for all the dwarf stars in APOGEE DR17 with period measurements in the background gray points, and the colored symbols show the genuinely young \halpha\ dwarf stars we have identified, with the colors and symbols as indicated in the legend in the top figure.
The ages are determined using \texttt{GPgyro}.
The bottom left plot shows the Li-age relation for solar twin stars \citep{Carlos2016}.
The ages in their sample is derived from isochrone fitting \citep{Nissen2015}, and the Li abundances are measured using HARPS spectra \citep{Mayor2003}.
Despite the differences in age derivation techniques and the instruments being used to derive abundances, the four genuinely young \halpha\ stars follow the typical Li-age trend of solar twins of intermediate ages.
One caveat is that these solar twins are near solar metallicity, and our sample has a metallicity of $\sim-0.5$ dex.
However, for their \halpha\ star sample, only upper limit for A(Li)$<$0.9 dex is detected, unlike the solar-mass stars in our sample.
As a result, this comparison suggests the Li content for the genuinely young \halpha\ stars likely have not been altered from mass-transfers from a stellar or sub-stellar companion.
The bottom right plot shows the [Al/Fe]-[Fe/H] relation for all the \halpha\ dwarf stars in APOGEE DR17, as shown in the background histogram for reference, and the white outlined colored symbols show the three genuinely young \halpha\ dwarf stars, with Al measurements missing for 03435417+1719250.
Interestingly enough, the stars with common abundance patterns (elevated Al and Li compared to their doppelgängers) all clustered around a metallicity of $\sim-0.5$ dex and age about 5 Gyr.
Their [Al/Fe]-[Fe/H] abundances also trace the outer edge of the distribution, where stars are not common.
However, since the star formation efficiency in Sagittarius is low, the $\alpha$ abundances, including Mg and Al, for stars in the Sagittarius core are much lower than those of our genuinely young \halpha\ dwarf stars.
As a result, we do not draw any conclusions given the small sample of our stars.
Hopefully, future follow-up for more young \halpha\ candidates will increase the sample size and aid our understanding of this population and formation pathway.

\subsection{One Possible Mass-Transfer Product}\label{subsec:merger}
\begin{figure*}
    \centering
    \includegraphics[width=\linewidth]{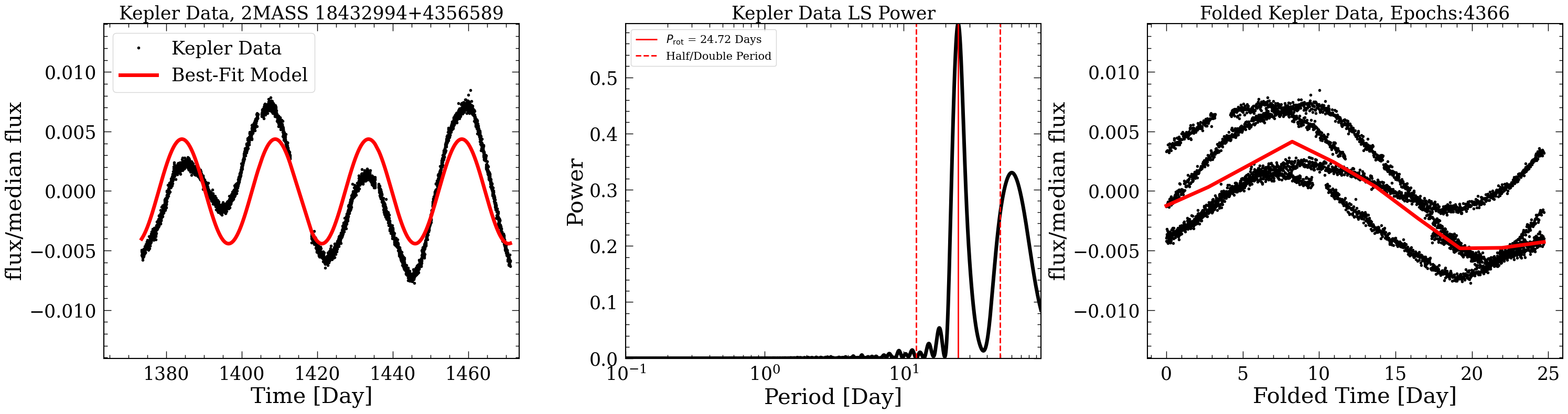}
    \includegraphics[width=\linewidth]{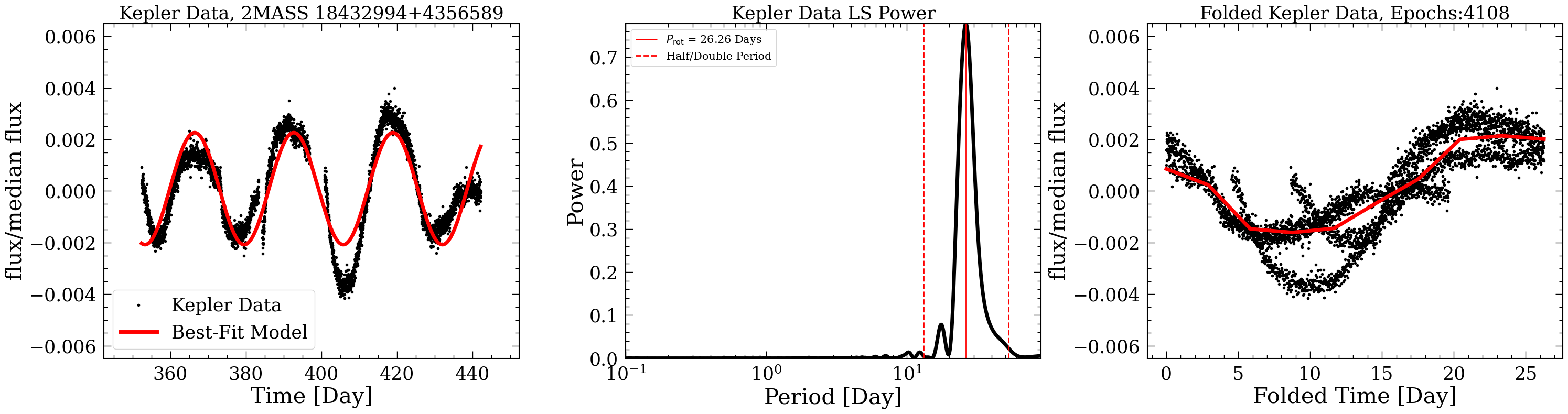}
    \includegraphics[width=0.24\linewidth]{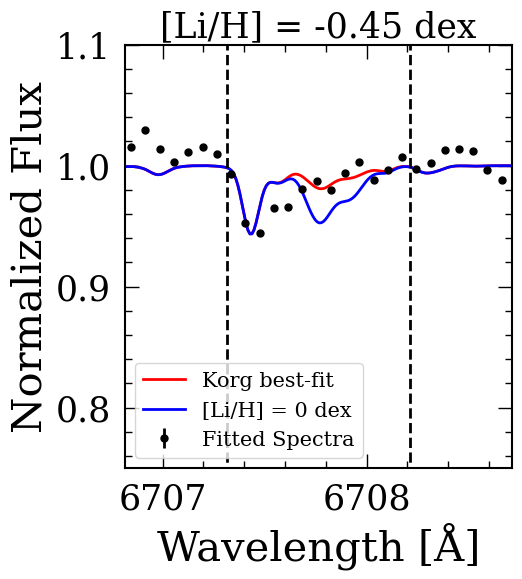}
    \includegraphics[width=0.24\linewidth]{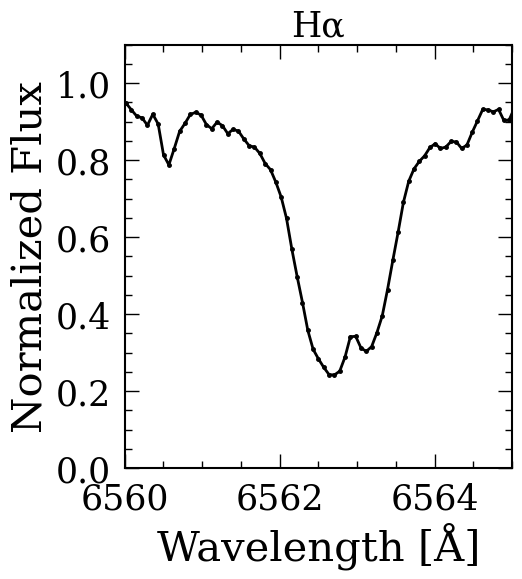}
    \includegraphics[width=0.48\linewidth]{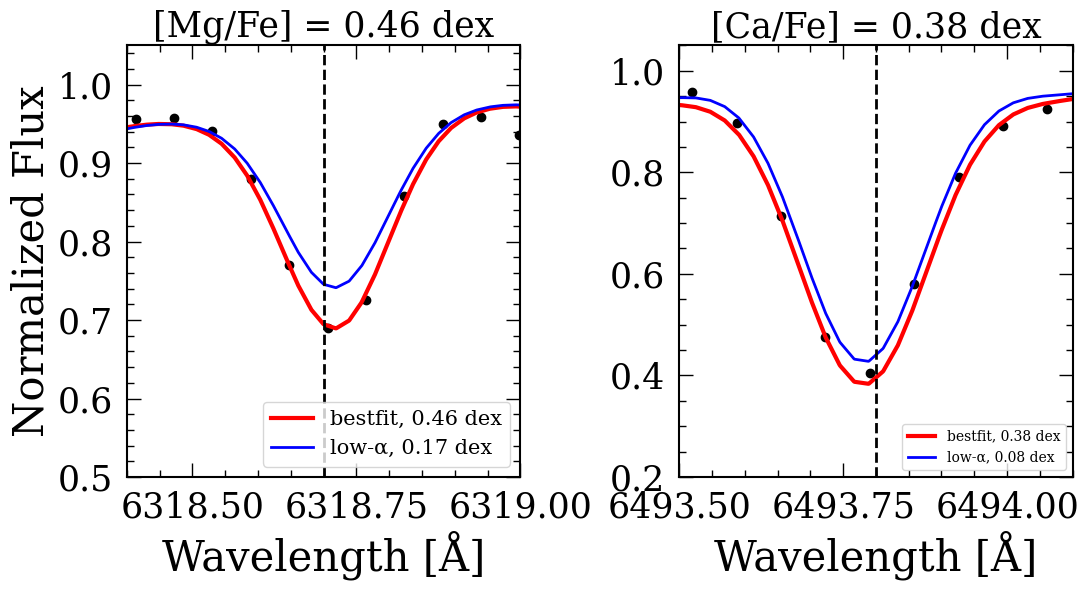}
    \includegraphics[width=0.43\linewidth]{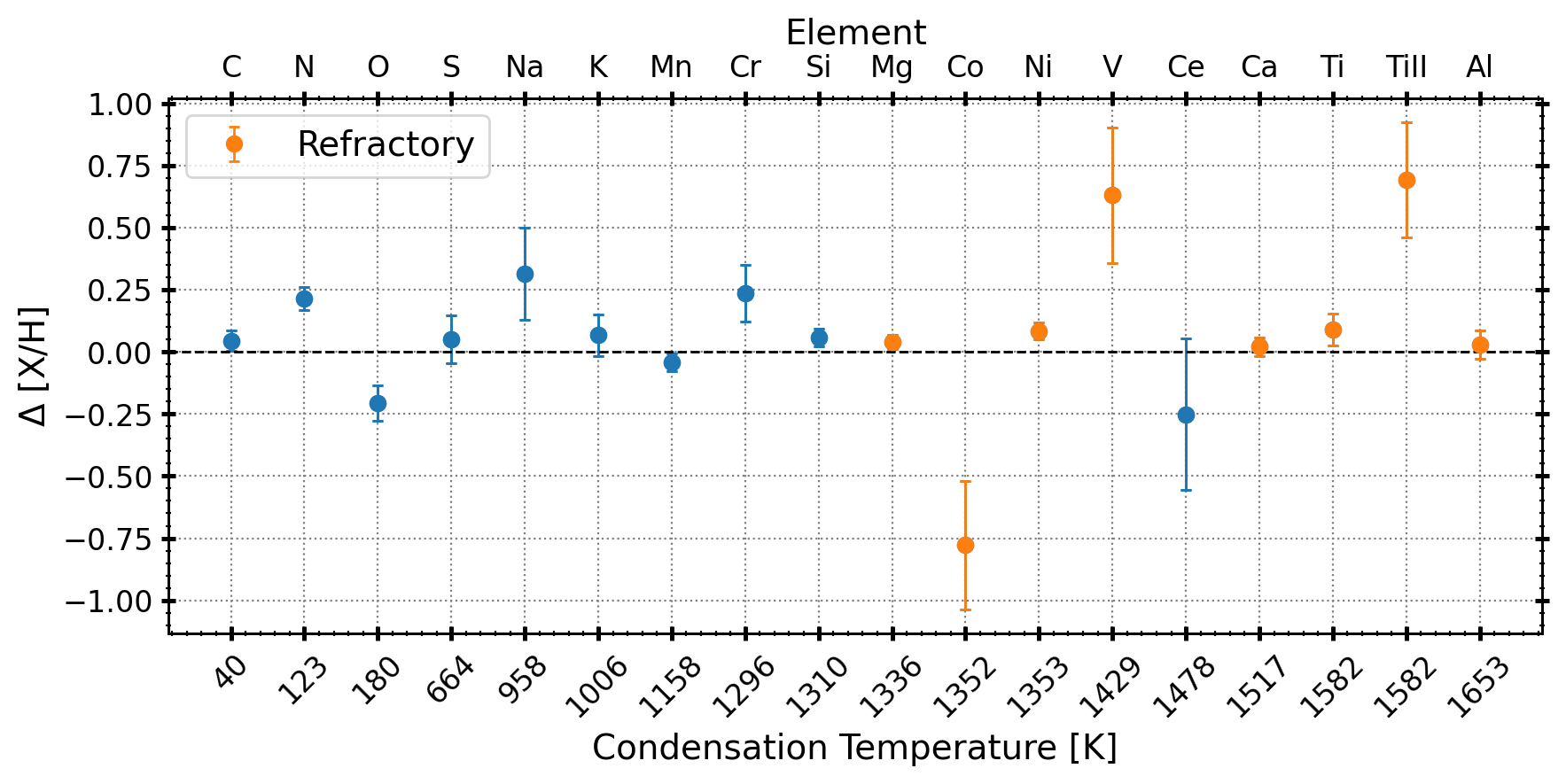} 
    \includegraphics[width=0.27\linewidth]{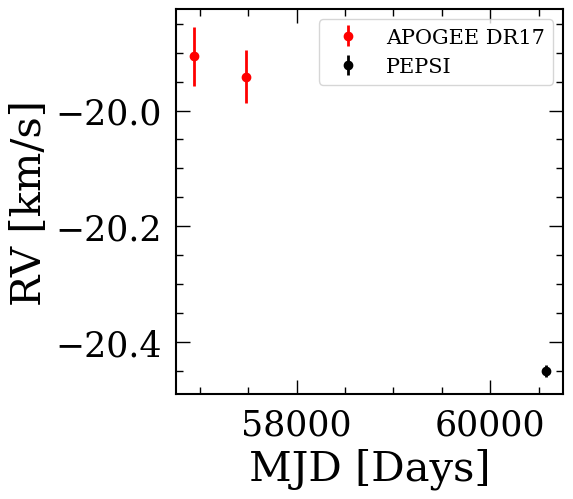} 
    \includegraphics[width=0.26\linewidth]{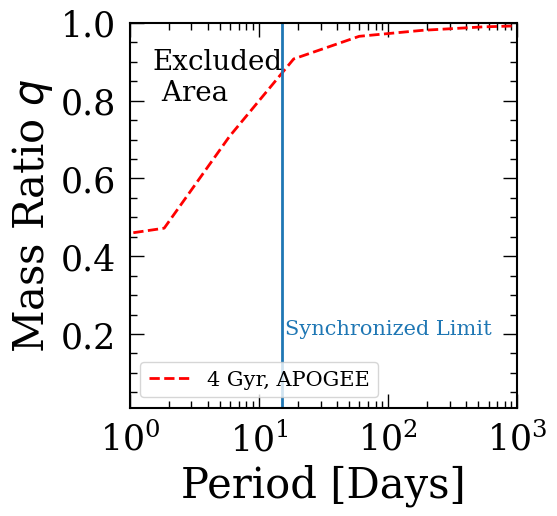} 
    \caption{Similar as Figure~\ref{fig:3} but for 18432994+4356589.
    We added the data around the H$\alpha$ line (see panel between the fits to Li and Mg in the middle row) to show the emission, indicating activity in this star. 
    This is an APOGEE--Kepler candidate star.
    Since this star is in the Kepler field, multiple sectors are available for rotation period measurements.
    We plotted two different Kepler sectors and their rotation analysis for this star in the top two rows.}
    \label{fig:7}
\end{figure*}

2MASS J18432994+4356589 shows a strong spot modulation signature based on Kepler data (see Figure~\ref{fig:7} top two rows), visible emission in the \halpha\ line core (see Figure~\ref{fig:7} middle row, middle panel).
The period for this star changes slightly between 24 and 26 days, matching what is expected for spot modulation due to differential rotation on the surface.
Both \halpha\ emission and strong spot modulation suggest its youth.
However, its \halpha\ nature suggests it should not have a strong activity signal.
Since this star has a \logg\ of 3.6, the activity signal is likely generated with a combination of the convection zone deepens and a relatively fast rotation.

We suspect this star has gone through stellar mass transfer or merger.
This star does not have a detectable Li line (see Figure~\ref{fig:7} middle left plot), suggesting either its Li has already been destroyed during its natural evolutionary process, or a mass transfer event happened and heated the star enough to burn its Li.
Since this star is evolved, it is entering the regime where Li is destroyed as the temperature at the base of the convective region becomes high enough when it ascends to the giant phase, so it is difficult to confirm whether it has gone through a mass transfer event.
Moreover, the limited RV measurements from APOGEE also do not exclude a binary companion.
However, since this star has a rotation period of over 20 days, it is likely not in a synchronized binary system. 
Future RV follow-ups will be able to provide better constraints on whether this star has a binary companion. 

The abundance comparison for this star shows strong disagreement between Ti and TiII measurements from APOGEE, suggesting possible incorrect \logg\ measurements, since by design, the ASPCAP pipeline \citep{APOGEEDR17} should return similar values for both.
%\textcolor{red}{Need to figure out exactly what this is, Jennifer?}
As a result, we are not surprised by the abundance variation compared to its doppelgänger stars.

Even though this star cannot be confirmed to be a genuinely young \halpha\ star, further follow-up observations can still yield interesting results on the formation pathway of \halpha\ stars with activity signatures indicating youth. 

\subsection{Formation pathways for genuinely young \halpha\ stars}\label{subsec:formation}
From these four stars, it is difficult to draw definitive conclusions on the formation pathways of genuinely young \halpha\ stars, but it is likely they formed in local, self-enriched environments before the enriched gas is dispersed in the Galaxy.
For example, the chemical evolution model described in \cite{Johnson2021} has predicted these stars, mostly formed in the outer disk, caused by highly variable SN Ia rate in the outer Galaxy due to sudden stellar migration of a sub-population of stars \citep{Dubay2024}.
These stars exhibit similar chemical abundance patterns in \alphafe\ and [Fe/H], in that they are the most metal-poor star for their age, with moderately elevated $\alpha$ ($\sim$0.1 dex).
The elevated Al could also indicate possible processes happening in globular clusters, as stars in globular clusters also exhibit elevated Al compared to predictions from chemical evolution models \citep[e.g.,][]{Sit2024, meszaros2020}.
However, it is worth pointing out there is a metallicity dependence in Al enhancements produced by H-burning Mg-Al chain \citep[e.g.,][]{Kobayashi2006, meszaros2020}, since our stars have [Fe/H] $>-$1.5 dex, the self-enrichment happening in globular clusters is likely not significant in the metallicity range of our stars.

\section{Conclusion \& Future Work}\label{sec:concl}
We performed high-resolution spectroscopic follow-up of 4 APOGEE--Kepler, 5 APOGEE--K2, and 4 GALAH--K2 young \halpha\ dwarf star candidates using PEPSI on LBT. 
Of the 13 observed stars, we have excluded 1 star with \halpha\ but low Mg measurement from APOGEE, indicating incorrect stellar parameters or abundance measurements.
This also means we are not able to confirm it belongs to the \halpha\ disk.
We also excluded 1 other star with a rotation period suggesting it could be similar age of a typical \halpha\ star.
Of the remaining 11 stars, 3 stars (2 from GALAH-K2 and 1 from APOGEE-K2) pass our additional tests on their young age, detectable lithium abundance, and low chance of a significant binary companion.

Combing with the one genuinely young \halpha\ dwarf star found in \cite{Lu2025}, we found:
\begin{itemize}
    \item Two out of three in GALAH have similar elevated Li abundances, $\Delta$A(Li)$\sim$0.5 dex, compared to their doppelgänger stars, suggesting they have not gone through binary mass transfer and their youth. 
    \item three out of four stars have similar elevated Al abundances, $\Delta$[Al/Fe]$\sim$0.1 dex, compared to their doppelgänger stars, suggesting their common chemical signature and thus, formation pathway. 
    \item three out of four stars are clustered at [Fe/H] = $-0.5$ dex, with ages close to 5 Gyr. 
    However, these stars are higher in $\alpha$ abundances (e.g., Mg and Al) compared to the stars in the core of Sagittarius, as the star formation efficiency in Sagittarius is low.
\end{itemize}

Our key result is that the young \halpha\ phenomenon is not restricted to asteroseismic mass measurements in giants. There are real cases where all of the available evidence points to youth; the stars that we have identified would not have been classified as old, or unusual, except relative to expectations for their high \alphafe\ ratio. 
It is also undoubtedly true that many apparently young candidates are the products of stellar interactions. 
Moving forward, we therefore believe that this should be treated as a mixed population, and effort should be made to distinguish between both of these interesting families of objects.
Future spectra and multi-epoch RV follow-ups should confirm more genuinely young $\alpha$ dwarf stars.
Moreover, asteroseismic constrain from the PLATO mission \citep{plato} could also put additional constrain on the ages of these stars. 
With more detections of genuinely young $\alpha$ stars, we will be able to better understand their formation pathway.

\section{Acknowledgments}
Y.L. would like to thank the helpful inputs from the OSU Galaxy Group and Stars Group.
Observations have benefited from the use of ALTA Center (alta.arcetri.inaf.it) forecasts performed with the Astro-Meso-Nh model. Initialization data of the ALTA automatic forecast system come from the General Circulation Model (HRES) of the European Centre for Medium Range Weather Forecasts.
This work has made use of data from the European Space Agency (ESA)
mission Gaia,\footnote{\url{https://www.cosmos.esa.int/gaia}} processed by
the Gaia Data Processing and Analysis Consortium (DPAC).\footnote{\url{https://www.cosmos.esa.int/web/gaia/dpac/consortium}} Funding
for the DPAC has been provided by national institutions, in particular
the institutions participating in the Gaia Multilateral Agreement.
This research also made use of public auxiliary data provided by ESA/Gaia/DPAC/CU5 and prepared by Carine Babusiaux. 
All the Kepler and K2 data used in this paper can be found in MAST: \dataset[10.17909/T9059R]{http://dx.doi.org/10.17909/T9059R} and \dataset[10.17909/T93W28]{http://dx.doi.org/10.17909/T93W28} \citep{KIC_MAST, K2_MAST}.
Parts of the text and analysis were assisted by OpenAI’s ChatGPT (version GPT-4.5, accessed July 2025) for drafting and editing support.
% exoplanet archieve
%This research has made use of the NASA Exoplanet Archive, which is operated by the California Institute of Technology, under contract with the National Aeronautics and Space Administration under the Exoplanet Exploration Program \citep{NEA}.
SB acknowledges support from the Australian Research Council under grant number DE240100150.
LA acknowledges support from the Swiss National Science Foundation (SNF; Project 200021L-231331) and the French Agence Nationale de la Recherche (ANR-24-CE93-0009-01) “PRIMA - PRobing the origIns of the Milky WAy’s oldest stars”. CM is supported by the NSF Astronomy and Astrophysics Fellowship award number AST-2401638.

% OSG
%This research was done using services provided by the OSG Consortium \citep{OSG1,OSG2}, which is supported by the National Science Foundation awards \#2030508 and \#1836650.

% SIMBAD, Vizier, ADS
This research has also made use of NASA's Astrophysics Data System, 
and the VizieR \citep{vizier} and SIMBAD \citep{simbad} databases, 
operated at CDS, Strasbourg, France.

%% To help institutions obtain information on the effectiveness of their 
%% telescopes the AAS Journals has created a group of keywords for telescope 
%% facilities.
%
%% Following the acknowledgments section, use the following syntax and the
%% \facility{} or \facilities{} macros to list the keywords of facilities used 
%% in the research for the paper.  Each keyword is check against the master 
%% list during copy editing.  Individual instruments can be provided in 
%% parentheses, after the keyword, but they are not verified.

\vspace{5mm}
\facilities{LBT, PEPSI, Gaia, Kepler, K2, APOGEE, GALAH}

%% Similar to \facility{}, there is the optional \software command to allow 
%% authors a place to specify which programs were used during the creation of 
%% the manuscript. Authors should list each code and include either a
%% citation or url to the code inside ()s when available.

\software{  \texttt{BACCHUS} \citep{BACCHUS},
\texttt{Astropy} \citep{astropy:2013, astropy:2018, astropy2022},
\texttt{lightkurve} \citep{lightkurve2018},
            \texttt{Matplotlib} \citep{matplotlib}, 
            \texttt{NumPy} \citep{Numpy}, 
            \texttt{Pandas} \citep{pandas}, 
            \texttt{gala} \citep{gala2017},
            \texttt{galpy} \citep{galpy},
            \texttt{seaborn} \citep{seaborn},
            \texttt{The Joker} \citep{Joker},
            \texttt{ChatGPT} \citep{openai2025chatgpt},
            \texttt{Korg} \citep{korg},
            \texttt{GPgyro} \citep{Lu2024},
            \texttt{STAREVOL} \citep{Amard2019}}

%% Appendix material should be preceded with a single \appendix command.
%% There should be a \section command for each appendix. Mark appendix
%% subsections with the same markup you use in the main body of the paper.

%% Each Appendix (indicated with \section) will be lettered A, B, C, etc.
%% The equation counter will reset when it encounters the \appendix
%% command and will number appendix equations (A1), (A2), etc. The
%% Figure and Table counter will not reset.

%\appendix

%\section{Appendix information}

\bibliography{sample631}{}
\bibliographystyle{aasjournal}

%% This command is needed to show the entire author+affiliation list when
%% the collaboration and author truncation commands are used.  It has to
%% go at the end of the manuscript.
%\allauthors

%% Include this line if you are using the \added, \replaced, \deleted
%% commands to see a summary list of all changes at the end of the article.
%\listofchanges

\end{document}